\documentclass[10pt,preprint]{aastex}
\usepackage{epsfig,lscape}
\slugcomment{KSUPT-03/3 \hspace{0.5truecm} June 2003}

\newcommand{\lap}{\lower.5ex\hbox{$\; \buildrel < \over \sim \;$}}
\newcommand{\gap}{\lower.5ex\hbox{$\; \buildrel > \over \sim \;$}}
\tighten
 
\begin{document}
\title{COBE-DMR-Normalized Dark Energy Cosmogony}
\author{Pia Mukherjee\altaffilmark{1,2}, A.~J.~Banday\altaffilmark{3}, 
        Alain Riazuelo\altaffilmark{4}, 
        Krzysztof M. G\'orski\altaffilmark{5,6}, 
        and Bharat Ratra\altaffilmark{1}}

\altaffiltext{1}{Department of Physics, Kansas State University, 116 Cardwell
                 Hall, Manhattan, KS 66506.}
\altaffiltext{2}{Present address: Department of Physics and Astronomy, 
                 University of Oklahoma, 440 W. Brooks Street, Norman, OK 
                 73019.}
\altaffiltext{3}{Max-Planck Institut f\"ur Astrophysik,
                 Karl-Schwarzschildstrasse 1, Garching D-85741, Germany.}
\altaffiltext{4}{Service de Physique Th\'eorique, CEA/DSM/SPhT, Unit\'e de
                 recherche associ\'ee au CNRS, CEA/Saclay F-91191 
                 Gif-sur-Yvette c\'edex, France.}
\altaffiltext{5}{Jet Propulsion Laboratory, California Institute of 
                 Technology, MS 169-327, 4800 Oak Grove Drive, 
                 Pasadena, CA 91109.}
\altaffiltext{6}{Warsaw University Observatory, Aleje Ujazdowskie 4, 
                 00-478 Warszawa, Poland.}

\begin{abstract}
Likelihood analyses of the COBE-DMR sky maps are used to determine the
normalization of the inverse-power-law-potential scalar field dark energy
model. Predictions of the DMR-normalized model are compared to various 
observations to constrain the allowed range of model parameters. Although
the derived constraints are restrictive, evolving dark energy density 
scalar field models remain an observationally-viable alternative to the 
constant cosmological constant model.
\end{abstract}

\section{Introduction}

Indications are that at the present epoch the universe is dominated by 
dark energy (see, e.g., Peebles \& Ratra 2003 for a review and 
Bennett et al. 2003a for a summary of the recent WMAP results).
Einstein's cosmological constant $\Lambda$ is the earliest example
of dark energy, and more recently scalar field models in which the 
energy density slowly decreases with time, and thus behaves like 
a time-variable $\Lambda$, have been the subject of much study
(see, e.g., Peebles 1984; Peebles \& Ratra 1988, 2003; 
Padmanabhan 2003).\footnote{
Recent discussions of dark energy models include Dvali \& Turner (2003), 
Wetterich (2003), Carroll, Hoffman, \& Trodden (2003), Saini, Padmanabhan,
\& Bridle (2003), Jain, Dev, \& Alcaniz (2003), Rosati (2003), Alam, Sahni, 
\& Starobinsky (2003), Munshi, Porciani, \& Wang (2003), Caldwell, 
Kamionkowski, \& Weinberg (2003), Klypin et al. (2003), Silva \& Bertolami 
(2003), Sen \& Sen (2003), and Singh, Sami, \& Dadhich (2003), through which 
the earlier literature may be accessed.}

In this paper we focus on a simple dark energy scalar field ($\phi$)
model. In this model the scalar field potential energy density 
$V(\phi)$ at low redshift is $\propto \phi^{-\alpha}$, with $\alpha > 0$
(see, e.g., Peebles \& Ratra 1988; Ratra \& Peebles 1988). 
Consistent with the observational indications, the cosmological
model is taken to be spatially flat (see, e.g., Podariu et al.~2001b; 
Durrer, Novosyadlyj, \& Apunevych 2003; Page et al.~2003; 
Melchiorri \& \"Odman 2003). 

Dark energy models, which assume an early epoch of inflation to generate
the needed initial energy density fluctuations, do not predict the 
amplitude of these fluctuations. At present, the most robust method to 
fix this amplitude, and hence the normalization of the model, is to compare 
model predictions of the large angular scale spatial anisotropy in the 
cosmic microwave background (CMB) radiation to what is observed. To this
end we compute the model predictions as a function of the parameter 
$\alpha$ and other cosmological parameters by following Brax, Martin,
\& Riazuelo (2000). We then determine the normalization amplitude by
comparing these predictions to the COBE-DMR CMB anisotropy measurements
(Bennett et al. 1996) by following G\'orski et al. (1998).

In linear perturbation theory, a scalar field is mathematically 
equivalent to a fluid with time-dependent equation of state parameter 
$w = p/\rho$ and speed of sound squared $c_s^2 = \dot p/\dot\rho$, where $p$ 
is the pressure, $\rho$ the energy density, and the dots denote time 
derivatives (see, e.g., Ratra 1991). The XCDM parametrization of this dark 
energy model approximates $w$ as a constant, which is accurate in the 
radiation and matter dominated epochs but not in the current, dark energy 
scalar field dominated epoch. This XCDM approximation leads to particularly
inaccurate predictions for the large angular scale CMB anisotropy.
We emphasize, however, that we do not work in the XCDM approximation,
rather we explicitly integrate the dark energy scalar field and other
equations of motion (Brax et al. 2000).

With a robust determination of the normalization of the inverse-power-law
potential dark energy scalar field it is possible to test the model and
constrain model parameter values by comparing other model predictions 
to observational data. In this paper we compare the predicted value, as a 
function of model parameters, of the rms linear mass fluctuation averaged 
over an 8 $h^{-1}$ Mpc sphere\footnote{
Here the Hubble constant $H_0 = 100 h$ km s$^{-1}$ Mpc$^{-1}$.},
$\delta M/M(8 h^{-1} {\rm Mpc})$ or $\sigma_8$, to observational estimates 
of this quantity. We also present qualitative comparisons of the
predicted fractional energy-density perturbation power spectrum $P(k)$
to various observations.

Such tests probe a combination of both the early universe physics of 
inflation and late universe dark energy physics.\footnote{
Another such test is the detection of the late integrated Sachs-Wolfe 
effect from the cross-correlation of various large scale structure maps 
with maps of CMB temperature anisotropy (see Boughn \& Crittenden 2003; 
Fosalba, Gazta\~naga, \& Castander 2003; Myers et al.~2003; Nolta et 
al.~2003; Scranton et al.~2003).}  
Other neoclassical 
cosmological tests are largely insensitive to the early universe 
physics of inflation and are hence more direct probes of dark energy.
Of special interest are tests based on gravitational lensing (see,
e.g., Ratra \& Quillen 1992; Waga \& Frieman 2000; Chae et al.~2002),
Type Ia supernovae redshift-apparent magnitude (see, e.g.,  Podariu
\& Ratra 2000; Waga \& Frieman 2000; Leibundgut 2002; Tonry et al.~2003), 
and redshift-angular size (see, e.g.,  Chen \& Ratra 2003a; Podariu 
et al.~2003; Zhu \& Fujimoto 2003, also see Daly \& Djorgovski 2003) data.

In $\S$ 2 we summarize the techniques we use to determine the DMR 
estimate of the CMB rms quadrupole moment anisotropy amplitude 
$Q_{\rm rms-PS}$ for the inverse-power-law potential dark energy 
scalar field model. Results for $Q_{\rm rms-PS}$ and large-scale 
structure statistics for the DMR-normalized models are given in 
$\S$ 3. These statistics are compared to current observational 
measurements in $\S$ 4. Our results are summarized in $\S$ 5.

\section{Summary of Computation}

We consider a scale-invariant primordial energy-density fluctuation
power spectrum (Harrison 1970; Peebles \& Yu 1970; Zel'dovich 1972),
as is generated by quantum fluctuations in a weakly coupled scalar
field during an early epoch of inflation (see, e.g., Fischler, Ratra,
\& Susskind 1985), and consistent with the observational indications 
(see, e.g., Spergel et al.~2003). More precisely, the fractional 
energy-density perturbation power spectrum we consider is 
\begin{equation} 
  P(k) = A \, k \, T^2 (k) ,
\end{equation}
where $k\, (0 < k < \infty)$ is the magnitude of the coordinate 
spatial wavenumber, $T(k)$ is the transfer function, and $A$ is 
the normalization amplitude defined in terms of the Bardeen potential.

The CMB fractional temperature perturbation, $\delta T/T$, is 
expressed as a function of angular position, $(\theta, \phi)$, on the sky
via the spherical harmonic decomposition,
\begin{equation}
  {\delta T \over T}(\theta , \phi) = \sum_{\ell=2}^\infty
       \sum_{m=-\ell}^\ell a_{\ell m} Y_{\ell m}(\theta , \phi) .
\end{equation}
The CMB spatial anisotropy in a Gaussian model\footnote{
Simple inflation models are Gaussian, with fluctuations generated 
in a weakly coupled field (see, e.g., Ratra 1985; Fischler et al.~1985),
consistent with observational indications (see, e.g., Mukherjee, Hobson,
\& Lasenby 2000; Park et al.~2001; Komatsu et al.~2002, 2003; Santos et 
al.~2002; De Troia et al. 2003).} 
can then be characterized by the angular perturbation spectrum 
$C_\ell$, defined in terms of the ensemble average, 
\begin{equation}
   \langle a_{\ell m} a_{\ell^\prime m^\prime}{}^* \rangle = 
      C_\ell \delta_{\ell\ell^\prime} \delta_{mm^\prime} ,
\end{equation}
where the $\delta$'s are Kronecker delta functions.

The $C_\ell$'s used here were computed using the Boltzmann transfer code
of Brax et al.~(2000). The computations here assume a standard recombination 
thermal history, and ignore the possibility of early reionization, tilt, 
gravity waves, or space curvature. While observed early reionization 
(Bennett et al.~2003a) does affect the $C_\ell$'s on the large angular 
scales of interest here, the effect is not large and an accurate quantitative 
estimate awaits a better understanding of structure formation. At present 
there is not much observational motivation for the inclusion of tilt, 
gravity waves, or space curvature. 

The dark energy scalar field model we  consider here is characterized 
by the potential energy density function
\begin{equation}
  V(\phi) = \kappa \phi^{-\alpha} ,
\end{equation}
where the constant $\kappa$ has dimensions of mass raised to the power
$\alpha + 4$ (Peebles \& Ratra 1988). In the limit where the exponent
$\alpha$ approaches zero, the scalar field energy density is equivalent
to a constant cosmological constant $\Lambda$. Here the exponent $\alpha
\geq 0$. 

We evaluate the CMB anisotropy angular spectra for a range of 
inverse-power-law scalar field potential exponent $\alpha$ 
spanning the interval between 0 and 8 in steps of unity, a range 
of the matter density parameter $\Omega_0$ spanning the interval 
between 0.1 and 1 in steps
of 0.1 and for $\Omega_0 = 0.05$, and for a variety of values of
 the hubble parameter $h$ 
and the baryonic-mass density parameter $\Omega_B$. The values of 
$h$ were selected to cover the range of ages consistent with current 
requirements ($t_0 =$  11 Gyr, 13 Gyr, or 17 Gyr, see, e.g., 
Peebles \& Ratra 2003), with $h$ as a function of $\Omega_0$ 
computed accordingly. The values of $\Omega_B$ were chosen to be 
consistent with current standard nucleosynthesis constraints 
($\Omega_B h^2 =$ 0.006, 0.014, or 0.022, see, e.g., Peebles \& Ratra 
2003). To render the problem tractable, $C_\ell$'s were determined 
for the central values of $t_0$ and $\Omega_B h^2$, and for the two 
combinations of these parameters which most perturb the $C_\ell$'s 
from those computed at the central values (i.e., for the smallest 
$t_0$ we used the smallest $\Omega_B h^2$, and for the largest $t_0$ 
we used the largest $\Omega_B h^2$). Specific parameter values are 
given in columns (1) and (2) of Tables 1--3, and representative 
CMB anisotropy power spectra can be seen in Figs.~1 and 2. 

The differences in the low-$\ell$ shapes of the $C_\ell$'s in Figs.~1 
and 2 are a consequence of the interplay between the ``usual" (fiducial 
CDM) Sachs-Wolfe term and the ``integrated" Sachs-Wolfe term in the 
expression for the CMB spatial anisotropy (see, e.g., Hu \& Dodelson 2002 
for a review). The relative importance of these terms is determined by the 
values of $\Omega_0$ and $\alpha$. (These terms are not easy to estimate
analytically, and we can not trust the inaccurate XCDM approximation and 
must instead rely on numerical computations because these terms depend in 
complicated ways on the way in which $w$ changes from its value during the 
matter dominated epoch to its value at zero redshift.) More precisely, in dark 
energy models the gravitational potential decays with time at late time, 
so at late time the gravitational potential and its time derivative are of 
opposite sign. Consequently there is some anticorrelation between the usual 
and integrated Sachs-Wolfe terms. This anticorrelation more significantly 
reduces the $C_\ell$'s at larger $\Omega_0$ (where the integrated term is 
not too large) and larger $\alpha$ (because the transition from matter 
dominance to dark energy dominance occurs earlier).

The normalization of the theoretical CMB anisotropy spectra are 
determined by comparing them to the DMR data using the likelihood 
analysis method of G\'orski (1994). In this paper we utilize the DMR 
four-year co-added 53 and 90 GHz sky maps in galactic coordinates.
We do not consider the DMR ecliptic coordinates data here; the small 
shifts in the inferred normalization amplitudes due to the small 
differences in the galactic- and ecliptic-coordinate maps are quantified 
in G\'orski et al.~(1998). The 53 and 90 GHz maps are coadded using 
inverse-noise-variance weights. The main changes relative to the analysis 
of Gorski et al.~(1998) are that we use the DMR maps in the HEALPix 
pixelisation, and excise pixels likely to be contaminated by bright 
foreground emission on and near the Galactic plane  (Banday et al.~1997)
using a cut rederived from the DIRBE 140 $\mu$m data. The DIRBE map was 
reconstructed in the HEALPix format using  the publicly available 
Calibrated Individual Observations (CIO) files together with the DIRBE 
Sky and Zodi Atlas (DSZA, both products being described in Hauser et 
al.~1998), allowing the construction of a full sky map corrected for the 
zodiacal emission according to the model of Kelsall et al. (1998). We 
consider $C_\ell$'s to multipole $\ell \leq 30$. See G\`orski et al.~(1998) 
for a more detailed description of the method.

As mentioned, we excise sky-map pixels on and near the Galactic plane
where foreground emission dominates the CMB. Following G\'orski et 
al.~(1998, also see G\'orski et al.~1995, 1996) we quantify the extent to
which residual high-latitude Galactic emission can modify our results 
in two ways: by performing all computations both including and excluding 
the observed sky quadrupole, and by performing all computations 
with and without correcting for emission correlated with the DIRBE 140 
$\mu$m sky map, itself fitted to the data for each model.\footnote{
See G\'orski et al.~(1998) for a more detailed discussion. More recent
discussions of foreground emissions may be accessed through Mukherjee
et al.~(2002, 2003b), Banday et al.~(2003), and Bennett et al.~(2003b).}

\section{Results}

\subsection{Results of $Q_{\rm rms-PS}$ fitting}

The results of the DMR likelihood analysis are summarized in Figs. 3--9
and Tables 1--3.

We have computed DMR data likelihood functions $L(Q_{\rm rms-PS}, \Omega_0,
\alpha)$ using three sets of $C_\ell$'s computed for values of $(\Omega_B
h^2, t_0) = $ (0.014, 13 Gyr), (0.006, 11 Gyr), and (0.022, 17 Gyr). We 
have also considered four ``different" DMR data sets, either accounting for 
or ignoring the correction for faint high-latitude foreground Galactic 
emission, and either including or excluding the quadrupole moment from
the analysis.
   
Variations in $\Omega_0$ and $\alpha$ have the largest effect on the 
inferred value of $Q_{\rm rms-PS}$ from the DMR data. Changes in $\Omega_B
h^2$, $t_0$, faint high-latitude Galactic emission treatment, and quadrupole
moment treatment, have much smaller effects, consistent with the findings 
of G\'orski et al.~(1998). These results determine what we have chosen 
to display in Figs.~3--9 and Tables 1--3. 

Two representative sets of likelihood functions $L(Q_{\rm rms-PS}, \Omega_0)$
are shown in Figs.~3 and 4. Figure 3 shows those derived while accounting 
for the correction for faint high-latitude foreground Galactic emission, and 
including the quadrupole moment in the analysis. Figure 4 shows the likelihood 
functions derived while ignoring the faint high-latitude foreground Galactic 
emission correction, and excluding the quadrupole moment from the analysis. 
These data sets represent two different ways of dealing with effects of 
foreground emission on the inferred $Q_{\rm rms-PS}$ normalization
(with the former resulting in a typically 5 \% lower 
$Q_{\rm rms-PS}$ value). These two data sets span the range of 
normalizations inferred from our analysis here.\footnote{
The ecliptic-frame sky maps result in slightly larger inferred 
$Q_{\rm rms-PS}$ normalization values than the galactic-frame sky
maps used here, see G\'orski et al.~(1998).}
The other two data sets we have used here, accounting for the faint 
high-latitude foreground emission while excluding the quadrupole and
not accounting for the faint high-latitude foreground emission while 
including the quadrupole, result in $Q_{\rm rms-PS}$ 
normalizations that are typically 1--2 \% above the smaller value
or below the larger value of the first two data sets. 

Tables 1--3 give the $Q_{\rm rms-PS}$ central values and 1 and 
2 $\sigma$ ranges, computed from the appropriate posterior probability
density distribution function assuming a uniform prior. These tables are
computed for the three cases with $\Omega_B h^2 = 0.014$ and $t_0 = 13$
Gyr (Table 1),  $\Omega_B h^2 = 0.006$ and $t_0 = 11$ Gyr (Table 2),  
and $\Omega_B h^2 = 0.022$ and $t_0 = 17$ Gyr (Table 3). We present
results for four representative values of $\alpha$ (= 0, 2, 4, and 6),
and a range of values of $\Omega_0$ spanning the interval from 0.1 to 1. 
$Q_{\rm rms-PS}$ values are given only for the two extreme data sets  
whose likelihood functions are shown in Figs.~3 and 4: the faint high-latitude
Galactic foreground emission corrected, quadrupole included case, and the 
Galactic foreground emission uncorrected quadrupole excluded case.

Figure 5 shows the ridge lines of maximum likelihood $Q_{\rm rms-PS}$
value, as a function of $\Omega_0$, for two different values of $\alpha$,
for all four data sets used in this paper. The upper panels of Fig.~7 show
the effect the small differences between data sets have on the conditional
(fixed $\Omega_0$, $\alpha$, $\Omega_B h^2$, and $t_0$) likelihood 
function for $Q_{\rm rms-PS}$. 

Tables 1--3 also illustrate the shift in the inferred $Q_{\rm rms-PS}$
normalization amplitudes due to changes in $t_0$ and $\Omega_B h^2$. 
Figure 6 shows the effects that varying $t_0$ and $\Omega_B h^2$ have on 
the ridge lines of maximum likelihood $Q_{\rm rms-PS}$ as a function of 
$\Omega_0$. The lower panels of Fig.~7 show the effects varying $t_0$ and 
$\Omega_B h^2$ on the conditional (fixed $\Omega_0$ and $\alpha$) likelihood 
functions for $Q_{\rm rms-PS}$ for a given data set. On the whole, 
for the CMB anisotropy spectra considered here, shifts in $t_0$ and 
$\Omega_B h^2$ have only a small effect on the inferred normalization 
amplitude.

Future WMAP data (see, e.g., Bennett et al.~2003a) should be able to
significantly constrain the dark energy scalar field model. Here we use
the DMR normalized dark energy model predictions to semi-quantitatively
constrain the cosmological parameters of this model. Since we do not 
attempt to determine precise constraints on these parameters in this paper,
we do not combine results from analyses of all DMR data sets, as G\'orski
et al.~(1998) did for the open model (Gott 1982; Ratra \& Peebles 1995), 
to determine the most robust estimate of the $Q_{\rm rms-PS}$ 
normalization amplitudes.

Figures 8 and 9 show marginal likelihood functions for some of the 
cosmological model parameter values and DMR data sets considered here.
Figure 8 shows the likelihood function $L(\Omega_0, \alpha)$ derived by
marginalizing $L(\Omega_0, \alpha, Q_{\rm rms-PS})$ over $Q_{\rm rms-PS}$
with a uniform prior. Figure 9 shows conditional (on slices of constant
$\alpha$ or $\Omega_0$) projections of the two-dimensional likelihood 
functions of Fig.~8. The effects of Galactic foreground correction and
 inclusion of the quadrupole are important. While the DMR data
 qualitatively indicate that
lower $\alpha$ and higher $\Omega_0$ values are mildly favored, these
data by themselves do not provide significant quantitative constraints
on cosmological model parameter values.

\subsection{Computation of Large-Scale Structure Statistics}

The fractional energy-density perturbation power spectrum $P(k)$ (eq.~[1])
was evaluated from a numerical integration of the linear perturbation 
theory equations of motion. Table 4 lists the $P(k)$ normalization amplitudes 
$A$ (eq.~[1]) for a range of $\Omega_0$ and $\alpha$ values.\footnote{
Numerical noise and other small effects contribute to the tiny artificial 
differences between the various $\Omega_0 = 1$ model predictions in this 
and other tables.}
The normalization amplitude is quite insensitive to the
values of $\Omega_B h^2$ and $t_0$ used. It is more sensitive to the 
choice of DMR data set, and Fig.~10 shows representative $P(k)$'s
normalized to $Q_{\rm rms-PS}$ obtained by averaging those from the 
foreground-emission-corrected quadrupole-included analysis and the 
foreground-emission-uncorrected quadrupole-excluded analysis. Figure 10
also shows recent observational determinations of the galaxy power spectrum
 on linear scales.
The predicted matter power spectrum $P(k) \propto (Q_{\rm rms-PS})^2$
so the error in the DMR normalization of $P(k)$ is approximately twice
that in the DMR determination of $Q_{\rm rms-PS}$ (which is discussed below).
As always, when comparing model predictions to observational measurements
it is important to account for this normalization uncertainty, rather than
basing conclusions about model viability solely on the ``central" 
normalization value.

The mean square linear mass fluctuation averaged over a sphere of 
coordinate radius $\bar\chi$ is
\begin{equation}
   \left\langle\left[{\delta M\over M} (\bar\chi) \right]^2\right\rangle 
   = {9\over 2\pi^2} \int^\infty_0 dk\, {k^2 P(k) \over (k\bar\chi)^6}
   \left[{\rm sin}(k\bar\chi) - k\bar\chi \, {\rm cos}(k\bar\chi)\right]^2 .
\end{equation}
Tables 5--7 list predictions for $(\delta M/M) (8 h^{-1} {\rm Mpc})$ or
$\sigma_8$, the square root of the above expression evaluated at 
$8 h^{-1} {\rm Mpc}$, for various values of $\alpha$ and $\Omega_0$.
In Tables 5--7 we list the mean of the central values obtained from 
the foreground-emission-corrected quadrupole-included analysis and the 
foreground-emission-uncorrected quadrupole-excluded analysis.

G\'orski et al.~(1998) show that the maximal 1 $\sigma$ fractional 
$Q_{\rm rms-PS}$ (and so $\delta M/M$) uncertainty from the DMR
data ranges between about 10 \% and 12 \%, depending on model parameter
values, and the maximal 2 $\sigma$ uncertainty ranges between about 16 \% 
and 19 \%. As we have not done a complete analysis of all possible 
DMR data sets here, we adopt 10 \% and 20 \% as our 1 and 2 $\sigma$ 
DMR normalization uncertainties for the dark energy scalar field models,
for the purpose of comparing model predictions to observational measurements.

\section{Observational Constraints on DMR-Normalized Models}

Given the dependence of the DMR likelihoods on the quadrupole moment of
the CMB anisotropy, the DMR data do not meaningfully exclude any part of 
the ($\Omega_0$, $\alpha$, $t_0$, $\Omega_B h^2$) parameter space for 
the dark energy scalar field model considered here. In this section we 
combine current observational constraints on cosmological parameters 
(as summarized in Peebles \& Ratra 2003; Chen \& Ratra 2003b; Bennett
et al.~2003a) with the DMR-normalized model predictions to place 
semi-quantitative constraints on the range of allowed model-parameter 
values. 

We focus on the parameter values and predictions of Tables 5--7. These
are computed for parameter values  $\Omega_B h^2 = 0.014$ and $t_0 = 13$
Gyr (Table 5),  $\Omega_B h^2 = 0.006$ and $t_0 = 11$ Gyr (Table 6),  
and $\Omega_B h^2 = 0.022$ and $t_0 = 17$ Gyr (Table 7), which span the
two standard deviation range of interest:
\begin{eqnarray}
   11\ {\rm Gyr} \leq & t_0 & \leq 17\ {\rm Gyr} , \\
   0.006 \leq & \Omega_B h^2 & \leq 0.022 ,
\end{eqnarray} 
with central values of $t_0 = 13$ Gyr and $\Omega_B h^2 = 0.014$ (as
summarized in Peebles \& Ratra 2003; see Carretta et al.~2000, Krauss
\& Chaboyer 2001, and Chaboyer \& Krauss 2002 for $t_0$, and Burles,
Nollett, \& Turner 2001 and Cyburt, Fields, \& Olive 2001 for $\Omega_B 
h^2$). The recent WMAP measurements
(Bennett et al.~2003a) lie in the center of the above $t_0$ range,
$13.3\ {\rm Gyr} \leq t_0 \leq 14.1\ {\rm Gyr}\ (2\ \sigma)$, and near the
upper end of the above $\Omega_B h^2$ range $0.021 \leq \Omega_B h^2 
\leq 0.024\  (2\ \sigma)$, in good agreement with standard Big Bang 
nucleosynthesis theory and primordial deuterium abundance observations.

The first columns in Tables 5--7 list $\Omega_0$. A recent analysis 
of pre-WMAP constraints on $\Omega_0$ from a combination of dynamical,
baryon fraction, power spectrum, weak lensing, and cluster abundance
measurements results in the two standard deviation range
\begin{equation}
   0.2 \lap \Omega_0 \lap 0.35 
\end{equation}
(Chen \& Ratra 2003b), in striking accord with the Bennett et al.~(2003a)
WMAP estimate of $0.19 \leq \Omega_0 \leq 0.35\ (2\ \sigma)$. This 
constraint on $\Omega_0$ significantly narrows the range of viable 
model-parameter values.

From an analysis of all available measurements of the Hubble parameter
prior to mid 1999, Gott et al.~(2001) find at two standard deviations
\begin{equation}
   0.6 \leq h \leq 0.74 ,
\end{equation}
with central value $h = 0.67$. This is in good agreement with the WMAP
estimate $0.65 \leq h \leq 0.79\ (2\ \sigma)$ with central value $h = 0.71$ 
(Bennett et al. 2003a). This constraint on $h$ also significantly narrows 
the range of viable model parameters.

For an estimate of the rms linear fractional mass fluctuations at
$8 h^{-1}$ Mpc we use, at 2 $\sigma$,
\begin{equation}
   0.72 \leq \sigma_8 \leq 1.16 , 
\end{equation}
with $\sigma_8 = 0.94$ as central value (Chen \& Ratra 2003b), with 
an additional 2 $\sigma$ DMR normalization uncertainty of 20 \% added 
in quadrature. 

Focussing on the numerical values of Tables 5--7, and combining the
preceding constraints, there remains a small range of parameter space 
that is observationally viable. In particular, models with $\Omega_0 
\sim 0.3$ and $\alpha$ ranging from 0 to $\sim 3$ with $\Omega_B h^2 = 
0.014$ and $t_0 = 13$ Gyr, are viable, while larger values of $\Omega_B 
h^2 = 0.022$ and $t_0 = 17$ Gyr narrow the range of $\alpha$ significantly, 
barely allowing a constant $\Lambda$ model with $\alpha = 0$. Dark energy
models that deviate significantly from Einstein's cosmological constant
are not favored by the data, i.e., the data favor $\alpha$ closer to 0.
A model with $\alpha = 3$, as is still allowed, requires new physics near
a characteristic energy scale $\sim 10^3$ GeV (Peebles \& Ratra 1988, 
2003), which should soon be probed by accelerator-based elementary 
particle physics experiments.

\section{Conclusion}

We have normalized the inverse-power-law-potential scalar field dark 
energy model by comparing the predicted CMB anisotropy in this model to
that measured by the DMR experiment. In addition to effects associated
with the DMR data --- see G\'orski et al. (1998) for a more detailed 
discussion --- the model normalization is sensitive to the values of
$\Omega_0$ and $\alpha$ (the exponent of the inverse-power-law potential)
but almost independent of the values of $\Omega_B$ and $h$.

The DMR data alone can not be used to constrain $\Omega_0$ and $\alpha$
over the ranges considered here. Current cosmographic observations, in 
conjunction with current large-scale structure observations compared to 
the predictions of the DMR-normalized dark energy scalar field model,
do however provide quite restrictive constraints on model parameter 
values. This is largely because $\sigma_8$ depends sensitively on the 
values of $\alpha$ and $\Omega_0$ (see, e.g., Douspis et al. 2002), 
more so than the positions of the peaks in the CMB anisotropy spectrum. 

As discussed in the previous section, there is a small range of model 
parameter values where evolving dark energy density models remain an observationally-viable alternative to a constant cosmological constant 
model, however, the data do not favor large deviations from the constant
$\Lambda$ model. It would therefore be of interest to examine more
closely the supergravity-inspired generalization of the 
inverse-power-law-potential scalar field dark energy model (Brax \& Martin
2000; Brax et al. 2000), which has a scalar field potential that results
in a slower evolution of the dark energy density, more like a constant
$\Lambda$.

It is possible that current or future WMAP (see, e.g., Bennett et 
al.~2003a) and other CMB anisotropy data will significantly circumscribe 
this option (see, e.g., Baccigalupi et al. 2002; Mukherjee et al. 2003a; 
Caldwell \& Doran 2003), although it should be noted that the constraints 
derived here are quite restrictive since $\sigma_8$ is very sensitive to 
the values of $\alpha$ and $\Omega_0$. Among other experiments, the 
proposed SNAP space mission to measure the redshift-magnitude relation 
out to a redshift of two should also provide interesting constraints on 
evolving dark energy (see, e.g., Podariu, Nugent, \& Ratra 2001a; 
Wang \& Lovelace 2001; Weller \& Albrecht 2002; Gerke \& Efstathiou 
2002; Eriksson \& Amanullah 2002; Rhodes et al.~2003). 

\bigskip

We acknowledge the advice and assistance of P.~Brax, J.~Lesgourgues, 
and J.~Martin and thank the referee, M.~Vogeley, for helpful comments. 
PM and BR acknowledge support from NSF CAREER grant 
AST-9875031. This work was partially carried out at the Jet Propulsion 
Laboratory of the California Institute of Technology, under a contract 
with the National Aeronautics and Space Administration.

\clearpage

\thispagestyle{empty}
\begin{landscape}
\begin{deluxetable}{lccccccccccccc}
\tablecolumns{13} 
\tablewidth{0pt}
\tablecaption{$Q_{\rm rms-PS}$ Values for the $t_0 = 13$ Gyr,  
$\Omega_B h^2 = 0.014$ Models\tablenotemark{a}}
\tablehead{
\colhead{} & \colhead{} & \multicolumn{2}{c}{$\alpha$=0} & \colhead{} &
\multicolumn{2}{c}{$\alpha$=2} & \colhead{} & \multicolumn{2}{c}{$\alpha$=4} & 
\colhead{} & \multicolumn{2}{c}{$\alpha$=6} \\
\cline{3-4} \cline{6-7} \cline{9-10} \cline{12-13} \\
\multicolumn{2}{l}{G.C.\tablenotemark{b} :} & \colhead{Yes} & \colhead{No} & 
\colhead{} & \colhead{Yes} & \colhead{No} & \colhead{} & \colhead{Yes} & \colhead{No} & \colhead{} & \colhead{Yes} & \colhead{No} \\
\multicolumn{2}{c}{$\ell_{\rm min}$:} & \colhead{2} & \colhead{3} & 
\colhead{} & \colhead{2} & \colhead{3} & \colhead{} & \colhead{2} & 
\colhead{3} & \colhead{} & \colhead{2} & \colhead{3} \\
\cline{3-4} \cline{6-7} \cline{9-10} \cline{12-13} \\
\colhead{$\Omega_0$} & \colhead{$h$} & \colhead{$Q_{\rm rms-PS}$} & \colhead{$Q_{\rm rms-PS}$} & \colhead{} & \colhead{$Q_{\rm rms-PS}$} & 
\colhead{$Q_{\rm rms-PS}$} & \colhead{} & \colhead{$Q_{\rm rms-PS}$} & 
\colhead{$Q_{\rm rms-PS}$} & \colhead{} & \colhead{$Q_{\rm rms-PS}$} &
\colhead{$Q_{\rm rms-PS}$} \\
\colhead{} & \colhead{} & \colhead{($\mu$K)} & \colhead{($\mu$K)} & 
\colhead{} & \colhead{($\mu$K)} & \colhead{($\mu$K)} & \colhead{} & 
\colhead{($\mu$K)} & \colhead{($\mu$K)} & \colhead{} & 
\colhead{($\mu$K)} & \colhead{($\mu$K)} \\
\colhead{(1)} & \colhead{(2)} & \colhead{(3)} & \colhead{(4)} & 
\colhead{} & \colhead{(5)} & \colhead{(6)} & \colhead{} & 
\colhead{(7)} & \colhead{(8)} & \colhead{} & 
\colhead{(9)} & \colhead{(10)}} 
\startdata
\vspace{1mm}
0.1 & 0.96 & $22.52^{24.12\:25.84}_{21.00\:19.56}$ & $23.88^{25.52\:27.36}_{22.28\:20.80}$ & & $24.76^{26.60\:28.56}_{23.04\:21.40}$ & $26.36^{28.24\:30.28}_{24.56\:22.84}$ & & $26.68^{28.68\:30.80}_{24.84\:23.08}$ & $28.44^{30.48\:32.64}_{26.52\:24.64}$ & & $28.44^{30.52\:32.76}_{26.44\:24.56}$ & $30.32^{32.36\:34.64}_{28.20\:26.24}$
\\
\vspace{1mm}
0.2 & 0.81 & $20.20^{21.60\:23.12}_{18.84\:17.56}$ & $21.36^{22.88\:24.40}_{19.96\:18.64}$ & & $22.52^{24.12\:25.92}_{20.96\:19.52}$ & $23.88^{25.56\:27.40}_{22.28\:20.80}$ & & $24.32^{26.12\:28.04}_{22.68\:21.04}$ & $25.84^{27.72\:29.72}_{24.12\:22.48}$ & & $25.76^{27.64\:29.68}_{23.96\:22.24}$ & $27.40^{29.32\:31.44}_{25.52\:23.76}$
\\
\vspace{1mm}
0.3 & 0.72 & $18.80^{20.08\:21.48}_{17.56\:16.36}$ & $19.84^{21.20\:22.64}_{18.60\:17.36}$ & & $20.56^{22.04\:23.60}_{19.20\:17.88}$ & $21.76^{23.28\:24.92}_{20.36\:19.00}$ & & $22.20^{23.80\:25.56}_{20.72\:19.28}$ & $23.56^{25.20\:27.00}_{22.00\:20.52}$ & & $23.44^{25.12\:26.96}_{21.84\:20.32}$ & $24.88^{26.60\:28.52}_{23.20\:21.64}$ \\
\vspace{1mm}
0.4 & 0.67 & $17.96^{19.20\:20.52}_{16.80\:15.68}$ & $18.96^{20.24\:21.60}_{17.76\:16.60}$ & & $18.84^{20.16\:21.60}_{17.60\:16.44}$ & $19.92^{21.28\:22.76}_{18.64\:17.44}$ & & $20.16^{21.56\:23.12}_{18.84\:17.56}$ & $21.32^{22.80\:24.40}_{19.96\:18.64}$ & & $21.16^{22.68\:24.28}_{19.76\:18.40}$ & $22.40^{23.96\:25.64}_{20.96\:19.56}$ \\ 
\vspace{1mm}
0.5 & 0.62 & $17.40^{18.60\:19.88}_{16.28\:15.24}$ & $18.36^{19.60\:20.92}_{17.24\:16.08}$ & & $17.52^{18.72\:20.00}_{16.40\:15.32}$ & $18.48^{19.72\:21.08}_{17.32\:16.20}$ & & $18.20^{19.44\:20.80}_{17.00\:15.88}$ & $19.20^{20.52\:21.92}_{18.00\:16.80}$ & & $18.92^{20.20\:21.64}_{17.68\:16.52}$ & $20.00^{21.36\:22.80}_{18.72\:17.48}$ \\
\vspace{1mm}
0.6 & 0.59 & $17.12^{18.28\:19.52}_{16.04\:15.00}$ & $18.04^{19.24\:20.56}_{16.92\:15.84}$ & & $16.48^{17.60\:18.84}_{15.44\:14.44}$ & $17.40^{18.52\:19.80}_{16.28\:15.24}$ & & $16.52^{17.64\:18.84}_{15.48\:14.44}$ & $17.40^{18.56\:19.84}_{16.32\:15.28}$ & & $16.80^{17.92\:19.16}_{15.72\:14.68}$ & $17.68^{18.88\:20.12}_{16.60\:15.52}$ \\ 
\vspace{1mm}
0.7 & 0.56 & $16.96^{18.12\:19.36}_{15.88\:14.84}$ & $17.88^{19.08\:20.36}_{16.76\:15.68}$ & & $15.96^{17.00\:18.20}_{14.92\:13.96}$ & $16.80^{17.92\:19.08}_{15.76\:14.76}$ & & $15.44^{16.48\:17.60}_{14.48\:13.52}$ & $16.24^{17.32\:18.48}_{15.24\:14.28}$ & & $15.16^{16.16\:17.24}_{14.20\:13.28}$ & $15.92^{16.96\:18.12}_{14.96\:14.00}$ \\
\vspace{1mm}
0.8 & 0.54 & $16.96^{18.08\:19.32}_{15.84\:14.80}$ & $17.88^{19.04\:20.32}_{16.72\:15.68}$ & & $15.88^{16.96\:18.08}_{14.88\:13.88}$ & $16.72^{17.80\:19.00}_{15.68\:14.68}$ & & $15.08^{16.08\:17.16}_{14.12\:13.20}$ & $15.88^{16.88\:18.04}_{14.88\:13.96}$ & & $14.48^{15.44\:16.48}_{13.60\:12.72}$ & $15.24^{16.24\:17.32}_{14.32\:13.40}$ \\
\vspace{1mm}
0.9 & 0.52 & $16.96^{18.12\:19.36}_{15.88\:14.84}$ & $17.92^{19.08\:20.36}_{16.76\:15.72}$ & & $16.32^{17.40\:18.60}_{15.28\:14.28}$ & $17.16^{18.32\:19.56}_{16.12\:15.08}$ & & $15.64^{16.72\:17.84}_{14.68\:13.72}$ & $16.48^{17.56\:18.72}_{15.48\:14.48}$ & & $15.24^{16.24\:17.32}_{14.28\:13.32}$ & $16.00^{17.08\:18.20}_{15.04\:14.08}$ \\
\vspace{1mm}
1 & 0.50 & $17.04^{18.20\:19.48}_{15.96\:14.92}$ & $18.00^{19.20\:20.48}_{16.88\:15.76}$ & & $17.08^{18.24\:19.52}_{16.00\:14.96}$ & $18.04^{19.20\:20.52}_{16.92\:15.80}$ & & $17.04^{18.20\:19.44}_{15.96\:14.88}$ & $17.96^{19.16\:20.44}_{16.84\:15.76}$ & & $17.04^{18.20\:19.48}_{15.96\:14.92}$ & $18.00^{19.20\:20.48}_{16.88\:15.76}$ \\ 
\enddata
\tablenotetext{a}{The tabulated $Q_{\rm rms-PS}$ values are determined 
from the conditional likelihood function at fixed $\Omega_0$ and $\alpha$. 
At each $\Omega_0$ and $\alpha$, the first of the five entries in each of 
columns (3)--(10) is the maximum likelihood value, the first (vertical) pair  
delimits the 68.3 \%  (1 $\sigma$) highest posterior density range, and  
the second (vertical) pair delimits the 95.5 \%  (2 $\sigma$) highest  
posterior density range.}
\tablenotetext{b}{Accounting for (Yes), or ignoring (No) the correction 
for faint high-latitude foreground Galactic emission.}
\end{deluxetable}
\end{landscape} 

\thispagestyle{empty}
\begin{landscape}
\begin{deluxetable}{lccccccccccccc}
\tablecolumns{13} 
\tablewidth{0pt}
\tablecaption{$Q_{\rm rms-PS}$ Values for the $t_0 = 11$ Gyr,  
$\Omega_B h^2 = 0.006$ Models\tablenotemark{a}}
\tablehead{
\colhead{} & \colhead{} & \multicolumn{2}{c}{$\alpha$=0} & \colhead{} &
\multicolumn{2}{c}{$\alpha$=2} & \colhead{} & \multicolumn{2}{c}{$\alpha$=4} & 
\colhead{} & \multicolumn{2}{c}{$\alpha$=6} \\
\cline{3-4} \cline{6-7} \cline{9-10} \cline{12-13} \\
\multicolumn{2}{l}{G.C.\tablenotemark{b} :} & \colhead{Yes} & \colhead{No} & 
\colhead{} & \colhead{Yes} & \colhead{No} & \colhead{} & \colhead{Yes} & \colhead{No} & \colhead{} & \colhead{Yes} & \colhead{No} \\
\multicolumn{2}{c}{$\ell_{\rm min}$:} & \colhead{2} & \colhead{3} & 
\colhead{} & \colhead{2} & \colhead{3} & \colhead{} & \colhead{2} & 
\colhead{3} & \colhead{} & \colhead{2} & \colhead{3} \\
\cline{3-4} \cline{6-7} \cline{9-10} \cline{12-13} \\
\colhead{$\Omega_0$} & \colhead{$h$} & \colhead{$Q_{\rm rms-PS}$} & \colhead{$Q_{\rm rms-PS}$} & \colhead{} & \colhead{$Q_{\rm rms-PS}$} & 
\colhead{$Q_{\rm rms-PS}$} & \colhead{} & \colhead{$Q_{\rm rms-PS}$} & 
\colhead{$Q_{\rm rms-PS}$} & \colhead{} & \colhead{$Q_{\rm rms-PS}$} &
\colhead{$Q_{\rm rms-PS}$} \\
\colhead{} & \colhead{} & \colhead{($\mu$K)} & \colhead{($\mu$K)} & 
\colhead{} & \colhead{($\mu$K)} & \colhead{($\mu$K)} & \colhead{} & 
\colhead{($\mu$K)} & \colhead{($\mu$K)} & \colhead{} & 
\colhead{($\mu$K)} & \colhead{($\mu$K)} \\
\colhead{(1)} & \colhead{(2)} & \colhead{(3)} & \colhead{(4)} & 
\colhead{} & \colhead{(5)} & \colhead{(6)} & \colhead{} & 
\colhead{(7)} & \colhead{(8)} & \colhead{} & 
\colhead{(9)} & \colhead{(10)}} 
\startdata
\vspace{1mm}
0.1 & 1.13 & $22.72^{24.36\:26.12}_{21.16\:19.68}$ & $24.12^{25.80\:27.64}_{22.52\:20.96}$ & & $24.96^{26.84\:28.84}_{23.24\:21.56}$ & $25.96^{28.52\:30.60}_{24.76\:23.04}$ & & $26.96^{28.96\:31.16}_{25.04\:23.24}$ & $28.72^{30.80\:32.96}_{26.76\:24.84}$ & & $28.76^{30.92\:33.16}_{26.76\:24.80}$ & $30.64^{32.76\:35.12}_{28.56\:26.52}$ \\ 
\vspace{1mm}
0.2 & 0.96 & $20.36^{21.80\:23.36}_{19.00\:17.72}$ & $21.56^{23.04\:24.68}_{20.16\:18.84}$ & & $22.68^{24.32\:26.12}_{21.12\:19.64}$ & $24.08^{25.80\:27.64}_{22.48\:20.92}$ & & $24.56^{26.36\:28.32}_{22.84\:21.24}$ & $26.12^{27.96\:30.00}_{24.32\:22.64}$ & & $26.00^{27.96\:30.04}_{24.20\:22.48}$ & $27.68^{29.68\:31.84}_{25.80\:24.04}$ \\
\vspace{1mm}
0.3 & 0.86 & $18.96^{20.28\:21.72}_{17.72\:16.52}$ & $20.04^{21.40\:22.88}_{18.76\:17.56}$ & & $20.76^{22.24\:23.84}_{19.36\:18.04}$ & $22.00^{23.52\:25.16}_{20.56\:19.16}$ & & $22.44^{24.04\:25.80}_{20.88\:19.44}$ & $23.80^{25.48\:27.28}_{22.24\:20.68}$ & & $23.64^{25.36\:27.24}_{22.04\:20.48}$ & $25.12^{26.88\:28.84}_{23.48\:21.84}$ \\
\vspace{1mm}
0.4 & 0.79 & $18.12^{19.36\:20.72}_{16.92\:15.80}$ & $19.12^{20.44\:21.84}_{17.92\:16.76}$ & & $19.04^{20.36\:21.80}_{17.76\:16.56}$ & $20.12^{21.52\:23.04}_{18.84\:17.60}$ & & $20.36^{21.80\:23.32}_{19.00\:17.68}$ & $21.52^{23.04\:24.64}_{20.16\:18.80}$ & & $21.36^{22.88\:24.52}_{19.96\:18.56}$ & $22.64^{24.20\:25.92}_{21.16\:19.76}$ \\ 
\vspace{1mm}
0.5 & 0.74 & $17.60^{18.80\:20.12}_{16.48\:15.36}$ & $18.56^{19.80\:21.16}_{17.40\:16.24}$ & & $17.68^{18.88\:20.24}_{16.52\:15.44}$ & $18.68^{19.92\:21.28}_{17.48\:16.32}$ & & $18.36^{19.64\:21.00}_{17.16\:16.04}$ & $19.40^{20.72\:22.16}_{18.16\:16.96}$ & & $19.12^{20.44\:21.88}_{17.88\:16.64}$ & $20.20^{21.60\:23.08}_{18.92\:17.68}$ \\
\vspace{1mm}
0.6 & 0.70 & $17.28^{18.44\:19.72}_{16.16\:15.08}$ & $18.24^{19.44\:20.76}_{17.08\:15.96}$ & & $16.60^{17.76\:19.00}_{15.56\:14.52}$ & $17.52^{18.72\:20.00}_{16.44\:15.36}$ & & $16.68^{17.80\:19.04}_{15.60\:14.56}$ & $17.60^{18.76\:20.04}_{16.48\:15.40}$ & & $16.96^{18.12\:19.36}_{15.88\:14.84}$ & $17.88^{19.08\:20.36}_{16.76\:15.68}$ \\ 
\vspace{1mm}
0.7 & 0.66 & $17.12^{18.24\:19.56}_{16.00\:14.96}$ & $18.04^{19.24\:20.56}_{16.92\:15.80}$ & & $16.08^{17.20\:18.36}_{15.08\:14.04}$ & $16.96^{18.08\:19.32}_{15.88\:14.88}$ & & $15.60^{16.64\:17.76}_{14.60\:13.64}$ & $16.40^{17.48\:18.68}_{15.36\:14.40}$ & & $15.28^{16.32\:17.44}_{14.32\:13.40}$ & $16.08^{17.16\:18.28}_{15.12\:14.12}$ \\
\vspace{1mm}
0.8 & 0.64 & $17.08^{18.24\:19.48}_{16.00\:14.92}$ & $18.00^{19.20\:20.52}_{16.88\:15.80}$ & & $16.04^{17.12\:18.28}_{15.00\:14.00}$ & $16.88^{18.00\:19.20}_{15.84\:14.80}$ & & $15.24^{16.24\:17.36}_{14.24\:13.32}$ & $16.00^{17.08\:18.20}_{15.04\:14.04}$ & & $14.64^{15.60\:16.68}_{13.72\:12.84}$ & $15.40^{16.40\:17.48}_{14.44\:13.52}$ \\
\vspace{1mm}
0.9 & 0.61 & $17.12^{18.32\:19.56}_{16.04\:15.00}$ & $18.08^{19.28\:20.60}_{16.92\:15.84}$ & & $16.44^{17.56\:18.76}_{15.40\:14.36}$ & $17.32^{18.48\:19.72}_{16.24\:15.20}$ & & $15.80^{16.88\:18.00}_{14.80\:13.80}$ & $16.64^{17.72\:18.92}_{15.60\:14.60}$ & & $15.36^{16.40\:17.48}_{14.40\:13.44}$ & $16.16^{17.20\:18.40}_{15.16\:14.20}$ \\
\vspace{1mm}
1 & 0.59 & $17.20^{18.40\:19.64}_{16.12\:15.04}$ & $18.16^{19.36\:20.68}_{17.00\:15.92}$ & & $17.24^{18.40\:19.68}_{16.12\:15.04}$ & $18.20^{19.40\:20.72}_{17.04\:15.96}$ & & $17.16^{18.36\:19.60}_{16.08\:15.00}$ & $18.12^{19.32\:20.64}_{16.96\:15.88}$ & & $17.20^{18.40\:19.64}_{16.12\:15.04}$ & $18.16^{19.36\:20.68}_{17.00\:15.92}$ \\ 
\enddata
\tablenotetext{a}{The tabulated $Q_{\rm rms-PS}$ values are determined 
from the conditional likelihood function at fixed $\Omega_0$ and $\alpha$. 
At each $\Omega_0$ and $\alpha$, the first of the five entries in each of 
columns (3)--(10) is the maximum likelihood value, the first (vertical) pair  
delimits the 68.3 \%  (1 $\sigma$) highest posterior density range, and  
the second (vertical) pair delimits the 95.5 \%  (2 $\sigma$) highest  
posterior density range.}
\tablenotetext{b}{Accounting for (Yes), or ignoring (No) the correction 
for faint high-latitude foreground Galactic emission.}
\end{deluxetable}
\end{landscape} 

\thispagestyle{empty}
\begin{landscape}
\begin{deluxetable}{lccccccccccccc}
\tablecolumns{13} 
\tablewidth{0pt}
\tablecaption{$Q_{\rm rms-PS}$ Values for the $t_0 = 17$ Gyr,  
$\Omega_B h^2 = 0.022$ Models\tablenotemark{a}}
\tablehead{
\colhead{} & \colhead{} & \multicolumn{2}{c}{$\alpha$=0} & \colhead{} &
\multicolumn{2}{c}{$\alpha$=2} & \colhead{} & \multicolumn{2}{c}{$\alpha$=4} & 
\colhead{} & \multicolumn{2}{c}{$\alpha$=6} \\
\cline{3-4} \cline{6-7} \cline{9-10} \cline{12-13} \\
\multicolumn{2}{l}{G.C.\tablenotemark{b} :} & \colhead{Yes} & \colhead{No} & 
\colhead{} & \colhead{Yes} & \colhead{No} & \colhead{} & \colhead{Yes} & \colhead{No} & \colhead{} & \colhead{Yes} & \colhead{No} \\
\multicolumn{2}{c}{$\ell_{\rm min}$:} & \colhead{2} & \colhead{3} & 
\colhead{} & \colhead{2} & \colhead{3} & \colhead{} & \colhead{2} & 
\colhead{3} & \colhead{} & \colhead{2} & \colhead{3} \\
\cline{3-4} \cline{6-7} \cline{9-10} \cline{12-13} \\
\colhead{$\Omega_0$} & \colhead{$h$} & \colhead{$Q_{\rm rms-PS}$} & \colhead{$Q_{\rm rms-PS}$} & \colhead{} & \colhead{$Q_{\rm rms-PS}$} & 
\colhead{$Q_{\rm rms-PS}$} & \colhead{} & \colhead{$Q_{\rm rms-PS}$} & 
\colhead{$Q_{\rm rms-PS}$} & \colhead{} & \colhead{$Q_{\rm rms-PS}$} &
\colhead{$Q_{\rm rms-PS}$} \\
\colhead{} & \colhead{} & \colhead{($\mu$K)} & \colhead{($\mu$K)} & 
\colhead{} & \colhead{($\mu$K)} & \colhead{($\mu$K)} & \colhead{} & 
\colhead{($\mu$K)} & \colhead{($\mu$K)} & \colhead{} & 
\colhead{($\mu$K)} & \colhead{($\mu$K)} \\
\colhead{(1)} & \colhead{(2)} & \colhead{(3)} & \colhead{(4)} & 
\colhead{} & \colhead{(5)} & \colhead{(6)} & \colhead{} & 
\colhead{(7)} & \colhead{(8)} & \colhead{} & 
\colhead{(9)} & \colhead{(10)}} 
\startdata
\vspace{1mm}
0.1 & 0.73 & $22.20^{23.76\:25.44}_{20.72\:19.28}$ & $23.52^{25.12\:26.88}_{21.96\:20.52}$ & & $24.48^{26.28\:28.20}_{22.80\:21.20}$ & $26.00^{27.88\:29.84}_{24.28\:22.60}$ & & $26.24^{28.20\:30.28}_{24.44\:22.72}$ & $27.92^{29.92\:32.04}_{26.04\:24.28}$ & & $27.84^{29.92\:32.08}_{25.92\:24.08}$ & $29.68^{31.72\:33.88}_{27.68\:25.76}$ \\ 
\vspace{1mm}
0.2 & 0.62 & $19.88^{21.24\:22.76}_{18.56\:17.32}$ & $21.00^{22.44\:23.96}_{19.68\:18.40}$ & & $22.28^{23.84\:25.60}_{20.76\:19.32}$ & $23.60^{25.24\:27.04}_{22.04\:20.56}$ & & $24.04^{25.08\:27.68}_{22.40\:20.84}$ & $25.56^{27.36\:29.28}_{23.84\:22.24}$ & & $25.36^{27.20\:29.20}_{23.60\:21.96}$ & $26.96^{28.88\:30.92}_{25.16\:23.44}$ \\
\vspace{1mm} 
0.3 & 0.55 & $18.60^{19.88\:21.24}_{17.40\:16.24}$ & $19.64^{20.96\:22.40}_{18.40\:17.20}$ & & $20.36^{21.76\:23.36}_{19.00\:17.72}$ & $21.56^{23.04\:24.60}_{20.16\:18.80}$ & & $21.96^{23.52\:25.24}_{20.48\:19.08}$ & $23.28^{24.92\:26.64}_{21.76\:20.28}$ & & $23.08^{24.76\:26.52}_{21.56\:20.04}$ & $24.52^{26.20\:28.08}_{22.88\:21.36}$ \\
\vspace{1mm}
0.4 & 0.51 & $17.76^{19.00\:20.28}_{16.64\:15.52}$ & $18.76^{20.00\:21.36}_{17.56\:16.44}$ & & $18.68^{20.00\:21.36}_{17.48\:16.28}$ & $19.76^{21.08\:22.52}_{18.48\:17.28}$ & & $19.92^{21.32\:22.84}_{18.64\:17.36}$ & $21.08^{22.52\:24.08}_{19.72\:18.44}$ & & $20.88^{22.36\:23.92}_{19.52\:18.16}$ & $22.12^{23.60\:25.24}_{20.68\:19.28}$ \\ 
\vspace{1mm}
0.5 & 0.48 & $17.20^{18.36\:19.64}_{16.12\:15.04}$ & $18.16^{19.36\:20.64}_{17.04\:15.92}$ & & $17.32^{18.48\:19.76}_{16.20\:15.12}$ & $18.24^{19.48\:20.80}_{17.12\:16.00}$ & & $18.00^{19.24\:20.56}_{16.84\:15.72}$ & $19.00^{20.28\:21.64}_{17.80\:16.64}$ & & $18.68^{19.92\:21.32}_{17.48\:16.28}$ & $19.72^{21.04\:22.44}_{18.48\:17.28}$ \\
\vspace{1mm}
0.6 & 0.45 & $16.96^{18.12\:19.36}_{15.92\:14.88}$ & $17.88^{19.08\:20.36}_{16.80\:15.72}$ & & $16.32^{17.40\:18.60}_{15.28\:14.28}$ & $17.20^{18.32\:19.56}_{16.12\:15.08}$ & & $16.36^{17.44\:18.68}_{15.32\:14.32}$ & $17.24^{18.36\:19.60}_{16.16\:15.12}$ & & $16.56^{17.68\:18.92}_{15.52\:14.52}$ & $17.48^{18.60\:19.84}_{16.36\:15.32}$ \\ 
\vspace{1mm}
0.7 & 0.43 & $16.76^{17.92\:19.12}_{15.72\:14.68}$ & $17.68^{18.84\:20.12}_{16.56\:15.52}$ & & $15.80^{16.88\:18.04}_{14.84\:13.84}$ & $16.64^{17.76\:18.92}_{15.64\:14.60}$ & & $15.28^{16.28\:17.36}_{14.32\:13.36}$ & $16.04^{17.08\:18.24}_{15.04\:14.12}$ & & $14.96^{15.92\:17.00}_{14.00\:13.12}$ & $15.72^{16.72\:17.84}_{14.76\:13.80}$ \\
\vspace{1mm}
0.8 & 0.41 & $16.80^{17.92\:19.12}_{15.72\:14.68}$ & $17.68^{18.84\:20.12}_{16.60\:15.52}$ & & $15.72^{16.80\:17.92}_{14.76\:13.76}$ & $16.56^{17.64\:18.84}_{15.52\:14.56}$ & & $14.96^{15.96\:17.04}_{14.04\:13.12}$ & $15.72^{16.76\:17.88}_{14.76\:13.84}$ & & $14.32^{15.28\:16.28}_{13.44\:12.60}$ & $15.08^{16.04\:17.08}_{14.16\:13.24}$ \\
\vspace{1mm}
0.9 & 0.40 & $16.84^{17.96\:19.20}_{15.76\:14.72}$ & $17.76^{18.92\:20.16}_{16.64\:15.56}$ & & $16.16^{17.24\:18.44}_{15.16\:14.16}$ & $17.00^{18.12\:19.36}_{15.96\:14.96}$ & & $15.52^{16.56\:17.68}_{14.56\:13.60}$ & $16.36^{17.40\:18.56}_{15.32\:14.36}$ & & $15.04^{16.00\:17.08}_{14.08\:13.16}$ & $15.80^{16.84\:17.92}_{14.84\:13.88}$ \\
\vspace{1mm}
1 & 0.38 & $16.84^{18.04\:19.28}_{15.84\:14.80}$ & $17.84^{19.00\:20.28}_{16.72\:15.64}$ & & $16.92^{18.08\:19.32}_{15.88\:14.84}$ & $17.84^{19.00\:20.28}_{16.72\:15.64}$ & & $16.92^{18.04\:19.28}_{15.84\:14.80}$ & $17.80^{19.00\:20.28}_{16.72\:15.64}$ & & $16.92^{18.08\:19.32}_{15.88\:14.84}$ & $17.84^{19.04\:20.32}_{16.76\:15.68}$ \\ 
\enddata
\tablenotetext{a}{The tabulated $Q_{\rm rms-PS}$ values are determined 
from the conditional likelihood function at fixed $\Omega_0$ and $\alpha$. 
At each $\Omega_0$ and $\alpha$, the first of the five entries in each of 
columns (3)--(10) is the maximum likelihood value, the first (vertical) pair  
delimits the 68.3 \%  (1 $\sigma$) highest posterior density range, and  
the second (vertical) pair delimits the 95.5 \%  (2 $\sigma$) highest  
posterior density range.}
\tablenotetext{b}{Accounting for (Yes), or ignoring (No) the correction 
for faint high-latitude foreground Galactic emission.}
\end{deluxetable}
\end{landscape}

\begin{deluxetable}{lcccccccccc}
\tablecolumns{9} 
\tablewidth{0pt}
\tablecaption{Fractional Energy-Density Perturbation Power Spectrum Normalization Factor $Ah^4$\tablenotemark{a}}
\tablehead{
\colhead{} & \colhead{} & \colhead{$\alpha$=0} & \colhead{} & \colhead{$\alpha$=2} & \colhead{} & \colhead{$\alpha$=4} & \colhead{} & \colhead{$\alpha$=6} \\
\colhead{$\Omega_0$} & \colhead{} & \colhead{$Ah^4$} & \colhead{} & \colhead{$Ah^4$} & \colhead{} & \colhead{$Ah^4$} & \colhead{} & \colhead{$Ah^4$}  \\
\colhead{} & \colhead{} & \colhead{($10^5$ Mpc$^4$)} & \colhead{} & \colhead{($10^5$ Mpc$^4$)} & \colhead{} & \colhead{($10^5$ Mpc$^4$)} & \colhead{} & \colhead{($10^5$ Mpc$^4$)} \\
\colhead{(1)} & \colhead{} & \colhead{(2)} & \colhead{} & \colhead{(3)} & \colhead{} & \colhead{(4)} & \colhead{} & \colhead{(5)}}
\startdata
0.1\ &\ & 153 (28.4) & & 307 (47.0) & & 525 (69.1) & & 802 (92.9) \\
0.2\ &\ & 62.9 (14.6) & & 125 (23.2) & & 208 (33.1) & & 303 (42.9) \\
0.3\ &\ & 36.7 (9.82) & & 63.2 (14.1) & & 101 (19.3) & & 141 (24.2) \\
0.4\ &\ & 24.8 (7.28) & & 34.3 (9.14) & & 50.3 (11.7) & & 66.8 (14.1)\\
0.5\ &\ & 18.0 (5.64) & & 19.9 (6.14) & & 25.0 (7.16) & & 30.8 (8.13)\\
0.6\ &\ & 13.8 (4.48) & & 12.4 (4.31) & & 12.9 (4.49) & & 13.9 (4.69)\\
0.7\ &\ & 11.0 (3.62)& & 8.62 (3.21) & & 7.56 (3.01) & & 6.92 (2.87) \\
0.8\ &\ & 8.99 (2.96) & & 6.89 (2.59) & & 5.51 (2.30) & & 4.55 (2.06) \\
0.9\ &\ & 7.46 (2.45) & & 6.29 (2.24) & & 5.28 (2.05) & & 4.58 (1.88) \\
1\   &\ & 6.29 (2.05) & & 6.31 (2.05) & & 6.27 (2.05) & & 6.29 (2.05) 
\enddata
\tablenotetext{a}{Normalized to the mean of the central $Q_{\rm rms-PS}$ 
in the foregrou\-nd-emissi\-on-corrected quadrupole-included and the 
foreground-emission-un\-correc\-ted quadrupole-excluded cases (normalized to 
$Q_{\rm rms-PS} = 10$ $\mu$K); scale as $(Q_{\rm rms-PS})^2$. These are 
for the $t_0 = 13$ Gyr and $\Omega_B h^2 = 0.014$ models and are insensitive 
to the values of these parameters.} 
\end{deluxetable}

\begin{deluxetable}{lcccccccccc}
\tablecolumns{10} 
\tablewidth{0pt}
\tablecaption{$(\delta M/M) (8 h^{-1} {\rm Mpc})$ for the $t_0 = 13$ Gyr, $\Omega_B h^2=0.014$ Models\tablenotemark{a}}
\tablehead{
\colhead{} & \colhead{} & \colhead{} & \colhead{$\alpha$=0} & \colhead{} & \colhead{$\alpha$=2} & \colhead{} & \colhead{$\alpha$=4} & \colhead{} & \colhead{$\alpha$=6} \\
\colhead{$\Omega_0$} & \colhead{$h$} & \colhead{} & \colhead{$\sigma_8$} & \colhead{} & \colhead{$\sigma_8$} & \colhead{} & \colhead{$\sigma_8$} & \colhead{} & \colhead{$\sigma_8$} \\
\colhead{(1)} & \colhead{(2)} & \colhead{} & \colhead{(3)} & \colhead{} & \colhead{(4)} & \colhead{} & \colhead{(5)} & \colhead{} & \colhead{(6)}}
\startdata
0.1 & 0.96 & \hspace{4mm} & 0.632 & \hspace{4mm} & 0.295 & \hspace{4mm} & 0.169 & \hspace{4mm} & 0.107 \\  
0.2 & 0.81 & & 0.888 & & 0.509 & & 0.332 & & 0.234 \\
0.3 & 0.72 & & 1.02 & & 0.670 & & 0.479 & & 0.363 \\
0.4 & 0.67 & & 1.10 & & 0.798 & & 0.617 & & 0.495 \\
0.5 & 0.62 & & 1.14 & & 0.906 & & 0.745 & & 0.633 \\
0.6 & 0.59 & & 1.17 & & 0.991 & & 0.863 & & 0.771 \\
0.7 & 0.56 & & 1.18 & & 1.06 & & 0.972 & & 0.905 \\
0.8 & 0.54 & & 1.18 & & 1.11 & & 1.06 & & 1.02 \\
0.9 & 0.52 & & 1.18 & & 1.15 & & 1.12 & & 1.11 \\
1   & 0.50 & & 1.17 & & 1.17 & & 1.17 & & 1.17 
\enddata
\tablenotetext{a}{The mean of the central values obtained from 
the fore\-ground-emission-corrected quadrupole-included analysis and the 
foreground-emission-uncorrected quadrupole-excluded analysis}
\end{deluxetable}

\begin{deluxetable}{lcccccccccc}
\tablecolumns{10} 
\tablewidth{0pt}
\tablecaption{$(\delta M/M) (8 h^{-1} {\rm Mpc})$ for the $t_0 = 11$ Gyr, $\Omega_B h^2=0.006$ Models\tablenotemark{a}}
\tablehead{
\colhead{} & \colhead{} & \colhead{} & \colhead{$\alpha$=0} & \colhead{} & \colhead{$\alpha$=2} & \colhead{} & \colhead{$\alpha$=4} & \colhead{} & \colhead{$\alpha$=6} \\
\colhead{$\Omega_0$} & \colhead{$h$} & \colhead{} & \colhead{$\sigma_8$} & \colhead{} & \colhead{$\sigma_8$} & \colhead{} & \colhead{$\sigma_8$} & \colhead{} & \colhead{$\sigma_8$} \\
\colhead{(1)} & \colhead{(2)} & \colhead{} & \colhead{(3)} & \colhead{} & \colhead{(4)} & \colhead{} & \colhead{(5)} & \colhead{} & \colhead{(6)}}
\startdata
0.1 & 1.13 & \hspace{4mm} & 0.964 & \hspace{4mm} & 0.477 & \hspace{4mm} & 0.283 & \hspace{4mm} & 0.185 \\  
0.2 & 0.96 & & 1.26 & & 0.746 & & 0.498 & & 0.356 \\
0.3 & 0.86 & & 1.41 & & 0.945 & & 0.685 & & 0.522 \\
0.4 & 0.79 & & 1.49 & & 1.10 & & 0.854 & & 0.689 \\
0.5 & 0.74 & & 1.52 & & 1.22 & & 1.01 & & 0.862 \\
0.6 & 0.70 & & 1.54 & & 1.31 & & 1.15 & & 1.03 \\
0.7 & 0.66 & & 1.54 & & 1.39 & & 1.28 & & 1.19 \\
0.8 & 0.64 & & 1.53 & & 1.44 & & 1.38 & & 1.33 \\
0.9 & 0.61 & & 1.51 & & 1.47 & & 1.44 & & 1.44 \\
1   & 0.59 & & 1.49 & & 1.49 & & 1.49 & & 1.49 
\enddata
\tablenotetext{a}{The mean of the central values obtained from 
the fore\-ground-emission-corrected quadrupole-included analysis and the 
foreground-emission-uncorrected quadrupole-excluded analysis.}
\end{deluxetable}

\begin{deluxetable}{lcccccccccc}
\tablecolumns{10} 
\tablewidth{0pt}
\tablecaption{$(\delta M/M) (8 h^{-1} {\rm Mpc})$ for the $t_0 = 17$ Gyr, $\Omega_B h^2=0.022$ Models\tablenotemark{a}}
\tablehead{
\colhead{} & \colhead{} & \colhead{} & \colhead{$\alpha$=0} & \colhead{} & \colhead{$\alpha$=2} & \colhead{} & \colhead{$\alpha$=4} & \colhead{} & \colhead{$\alpha$=6} \\
\colhead{$\Omega_0$} & \colhead{$h$} & \colhead{} & \colhead{$\sigma_8$} & \colhead{} & \colhead{$\sigma_8$} & \colhead{} & \colhead{$\sigma_8$} & \colhead{} & \colhead{$\sigma_8$} \\
\colhead{(1)} & \colhead{(2)} & \colhead{} & \colhead{(3)} & \colhead{} & \colhead{(4)} & \colhead{} & \colhead{(5)} & \colhead{} & \colhead{(6)}}
\startdata
0.1 & 0.73 & \hspace{4mm} & 0.273 & \hspace{4mm} & 0.107 & \hspace{4mm} & 0.0540 & \hspace{4mm} & 0.0316 \\  
0.2 & 0.62 & & 0.453 & & 0.241 & & 0.149 & & 0.101 \\
0.3 & 0.55 & & 0.562 & & 0.352 & & 0.244 & & 0.179 \\
0.4 & 0.51 & & 0.631 & & 0.445 & & 0.336 & & 0.265 \\
0.5 & 0.48 & & 0.675 & & 0.523 & & 0.425 & & 0.356 \\
0.6 & 0.45 & & 0.707 & & 0.589 & & 0.509 & & 0.450 \\
0.7 & 0.43 & & 0.724 & & 0.645 & & 0.588 & & 0.544 \\
0.8 & 0.41 & & 0.740 & & 0.689 & & 0.655 & & 0.629 \\
0.9 & 0.40 & & 0.748 & & 0.725 & & 0.709 & & 0.698 \\
1   & 0.38 & & 0.751 & & 0.751 & & 0.750 & & 0.751 
\enddata
\tablenotetext{a}{The mean of the central values obtained from 
the fore\-ground-emission-corrected quadrupole-included analysis and the 
foreground-emission-uncorrected quadrupole-excluded analysis.}
\end{deluxetable}

\begin{figure}
\centerline{\epsfig{file=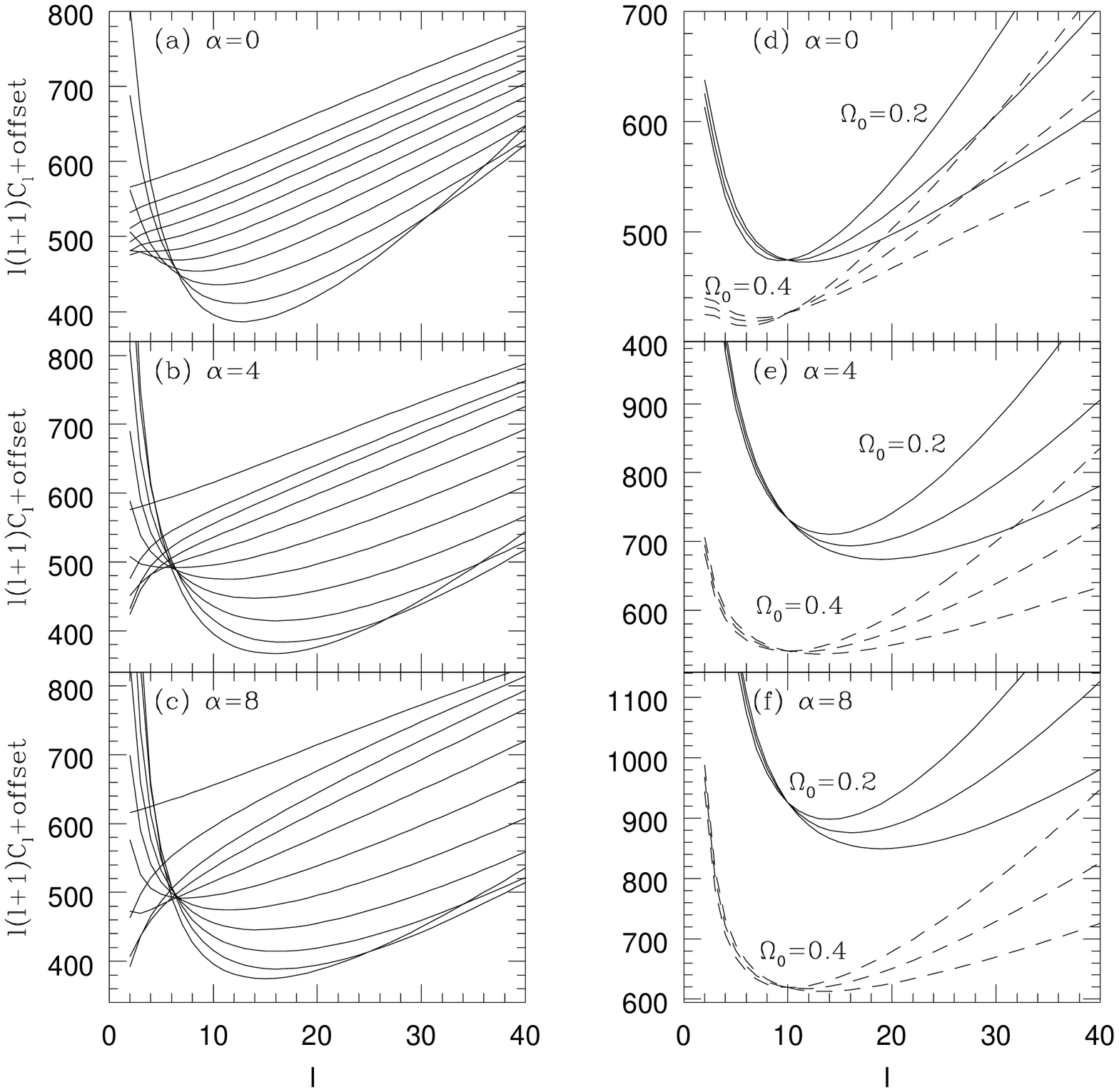,width=16.5cm}}
\caption{CMB anisotropy multipole coefficients. Panels $a)$ and $d)$ in the 
first row, $b)$ and $e)$ in the second row, and $c)$ and $f)$ in the third row
correspond to models with $\alpha$ = 0, 4, and 8, respectively. Panels
$a)$, $b)$, and $c)$ in the left column show multipoles for a model with
$t_0$ = 13 Gyr and $\Omega_B h^2$ = 0.014. The coefficients are normalized 
relative to the $C_9$ amplitude, and different values of $\Omega_0$ are 
offset from each other to aid visualization. In each of these panels, at $\ell
= 9$, the 11 curves, in ascending order, correspond to models with 
$\Omega_0 = 0.05, 0.1, 0.2, \dots, 0.9, 1.0$. The three panels in the
right column show models with $\Omega_0$ = 0.2 (solid curves) and 
0.4 (dashed curves). Each set of three curves in these panels correspond,
in ascending order on the right vertical axis of each panel, to models
with parameter values ($t_0$, $\Omega_B h^2$) = (11 Gyr, 0.006), 
(13 Gyr, 0.014), and (17 Gyr, 0.022), respectively.}
\label{f1}
\end{figure} 

\begin{figure}
\centerline{\epsfig{file=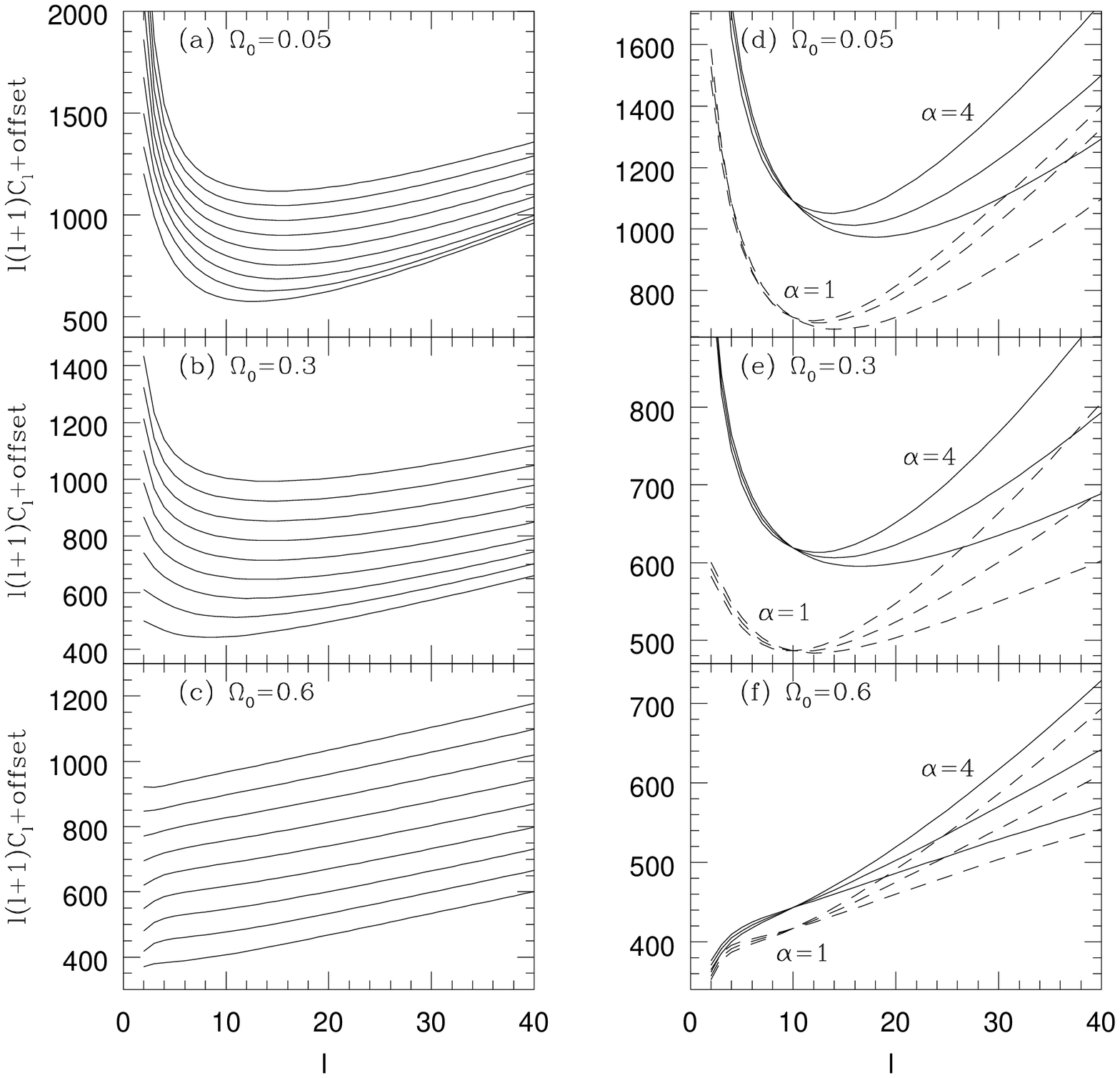,width=16.5cm}}
\caption{CMB anisotropy multipole coefficients. Panels $a)$ and $d)$ in the 
first row, $b)$ and $e)$ in the second row, and $c)$ and $f)$ in the third row
correspond to models with $\Omega_0$ = 0.05, 0.3, and 0.6, respectively. 
Panels $a)$, $b)$, and $c)$ in the left column show multipoles for a model 
with $t_0$ = 13 Gyr and $\Omega_B h^2$ = 0.014. The coefficients are 
normalized relative to the $C_9$ amplitude, and different values of $\alpha$ 
are offset from each other to aid visualization. In each of these panels, 
at $\ell = 9$, the 9 curves, in ascending order, correspond to models with 
$\alpha = 0, 1, 2, \dots, 7, 8$. The three panels in the
right column show models with $\alpha$ = 4 (solid curves) and 
1 (dashed curves). Each set of three curves in these panels correspond,
in ascending order on the right vertical axis of each panel, to models
with parameter values ($t_0$, $\Omega_B h^2$) = (11 Gyr, 0.006), 
(13 Gyr, 0.014), and (17 Gyr, 0.022), respectively.}
\label{f2}
\end{figure}

\begin{figure}
\centerline{\epsfig{file=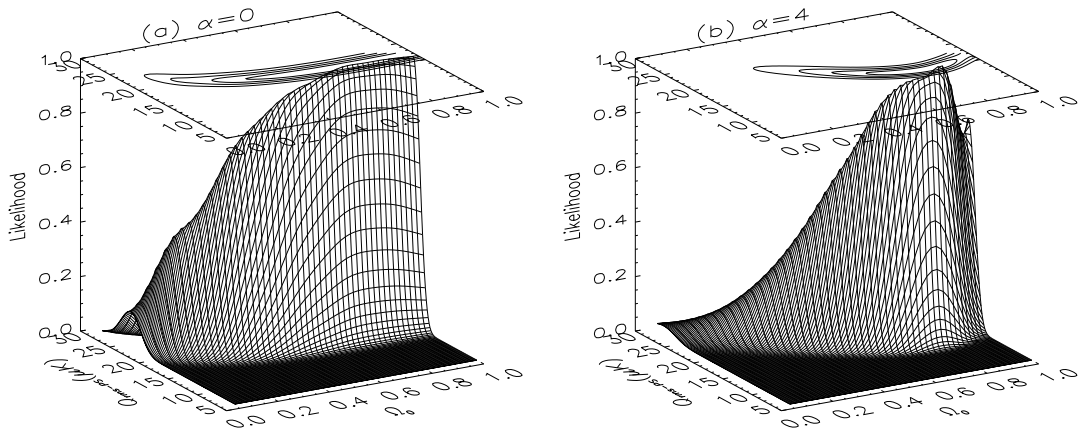,width=16.5cm}}
\caption{Likelihood functions $L(Q_{\rm rms-PS}, \Omega_0)$ (arbitrarily 
normalized to unity at the highest peak) derived from a simultaneous 
analysis of the DMR 53 and 90 GHz galactic-frame data, corrected for faint 
high-latitude foreground Galactic emission, and including the quadrupole 
moment in the analysis. These are for the model with $t_0$ = 13 Gyr and 
$\Omega_B h^2$ = 0.014. Panel $a)$ is for $\alpha = 0$ and panel $b)$ for 
$\alpha = 4$.}
\label{f3}
\end{figure}

\begin{figure}
\centerline{\epsfig{file=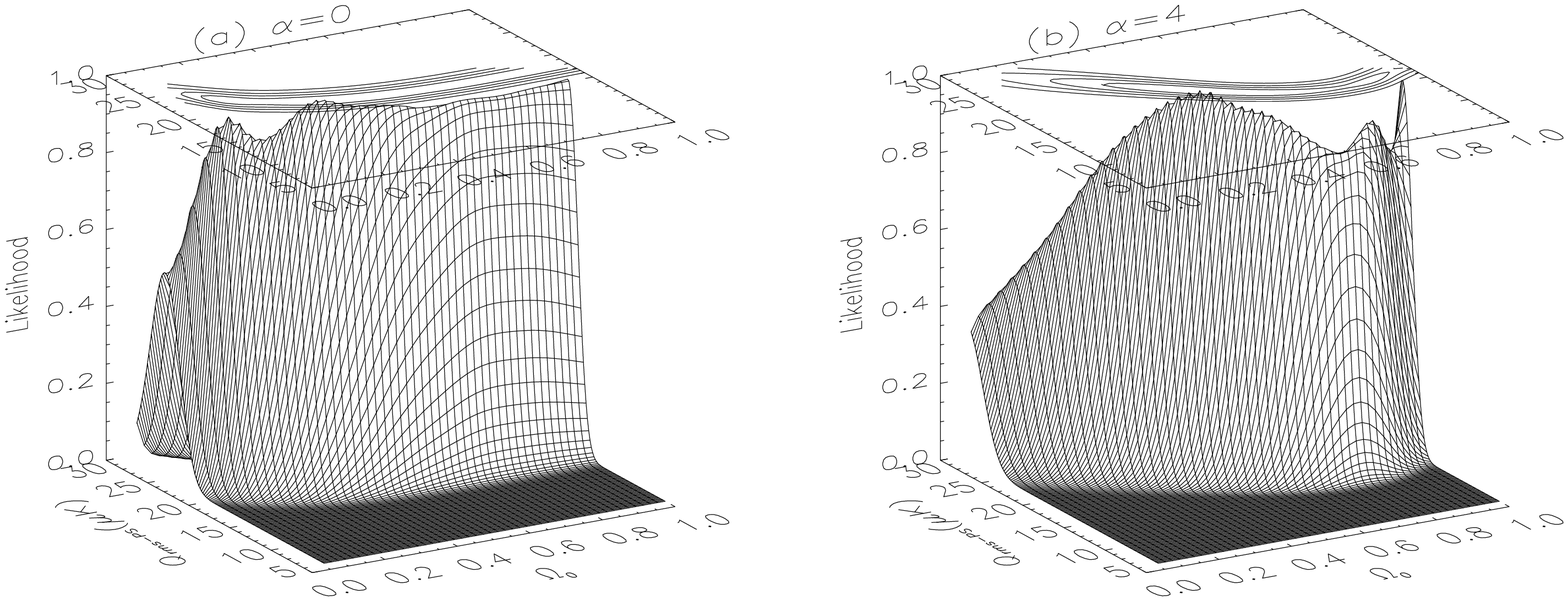,width=16.5cm}}
\caption{Likelihood functions $L(Q_{\rm rms-PS}, \Omega_0)$ (arbitrarily 
normalized to unity at the highest peak) derived from a simultaneous 
analysis of the DMR 53 and 90 GHz galactic-frame data, ignoring the correction
for faint high-latitude foreground Galactic emission, and excluding the 
quadrupole moment from the analysis. These are for the model with 
$t_0$ = 13 Gyr and $\Omega_B h^2$ = 0.014. Panel $a)$ is for $\alpha = 0$ 
and panel $b)$ for $\alpha = 4$.}
\label{f4}
\end{figure}

\begin{figure}
\centerline{\epsfig{file=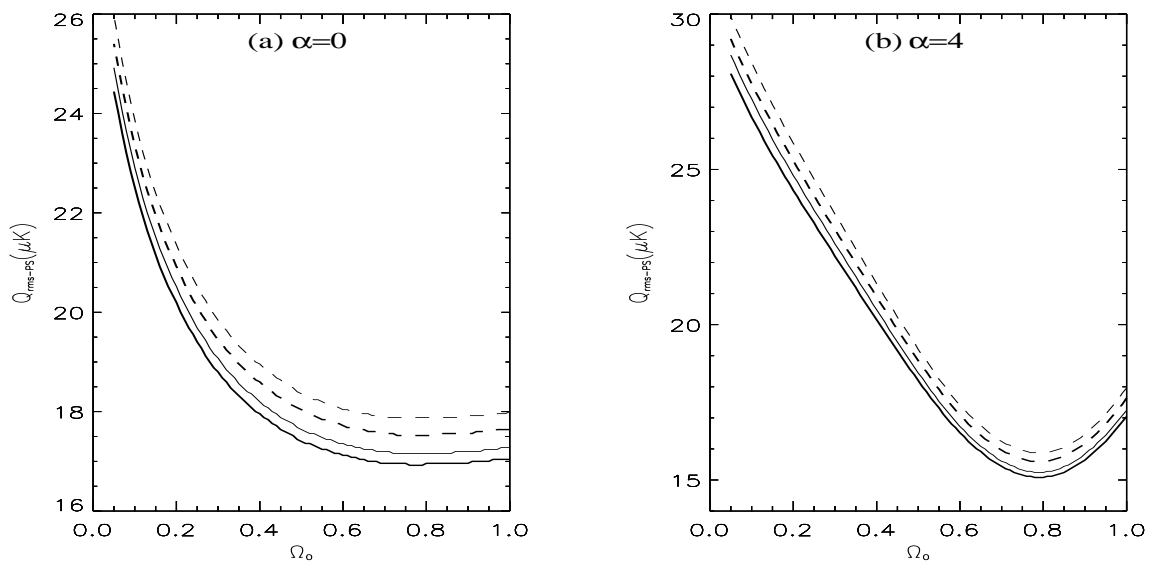,width=16.5cm, height=8cm}}
\caption{Ridge lines of the maximum likelihood $Q_{\rm rms-PS}$ value as 
a function of $\Omega_0$, for $a)$ $\alpha = 0$ and $b)$
$\alpha = 4$. These are for the model with $t_0$ = 13 
Gyr and $\Omega_B h^2$ = 0.014. Four curves are shown in each panel: 
solid (dashed) curves correspond to accounting for (ignoring) the faint 
high-latitude foreground Galactic emission correction, while heavy 
(light) curves include (exclude) the quadrupole moment in the analysis.}
\label{f5}
\end{figure}

\begin{figure}
\centerline{\epsfig{file=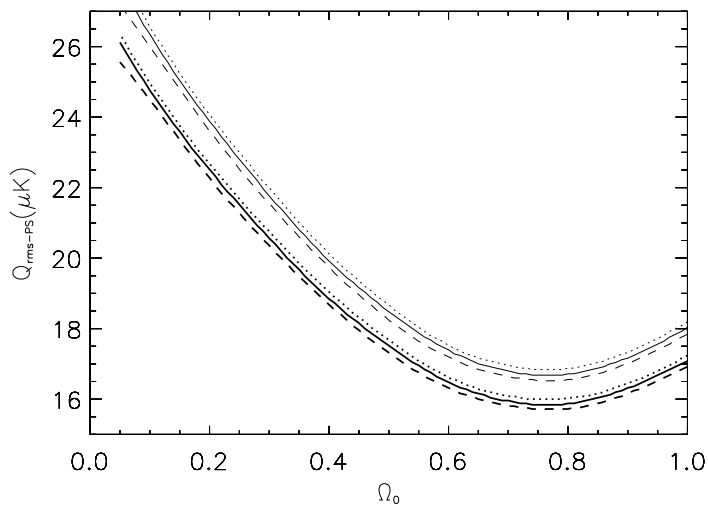,width=10.5cm}}
\caption{Ridge lines of the maximum likelihood $Q_{\rm rms-PS}$
value as a function of $\Omega_0$, for $\alpha = 2$. Six curves
are shown. These correspond to parameter values ($t_0$, $\Omega_B h^2$) = 
(11 Gyr, 0.006) (dotted curves), (13 Gyr, 0.014) (solid curves), and 
(17 Gyr, 0.022) (dashed curves). Heavy curves are determined by accounting 
for faint high-latitude foreground Galactic emission and including the 
quadrupole moment in the analysis, while light curves ignore the correction
for faint high-latitude foreground Galactic emission and exclude the 
quadrupole moment from the analysis.}
\label{f6}
\end{figure}

\begin{figure}
\centerline{\epsfig{file=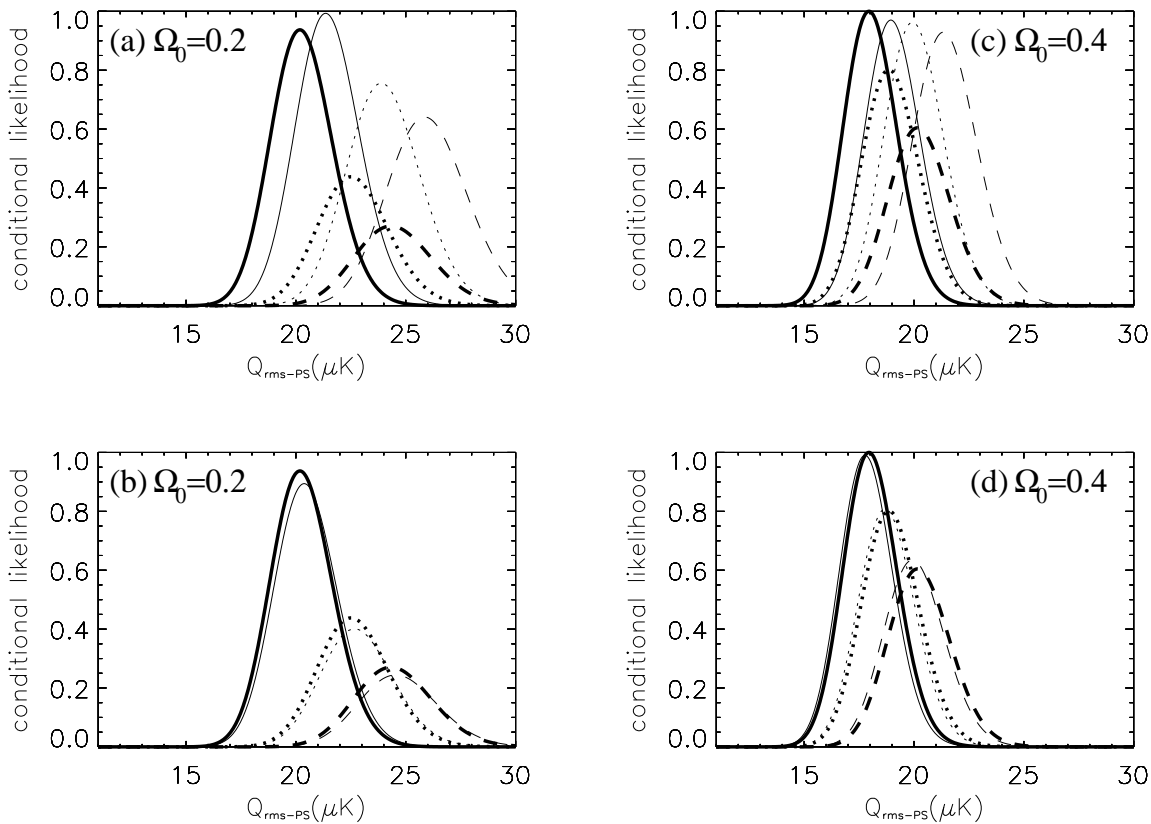,width=16.5cm}}
\caption{Conditional likelihood densities for $Q_{\rm rms-PS}$, derived 
from $L(Q_{\rm rms-PS}, \Omega_0, \alpha)$ (which are normalized to be 
unity at the peak, for each DMR data set and 
set of model-parameter values). Left column panels $a)$ and $b)$
are for $\Omega_0 = 0.2$ and right column panels $c)$ and $d)$
correspond to $\Omega_0 = 0.4$. Six curves are shown in each panel.
Solid, dotted, and dashed curves correspond to $\alpha$ = 0, 2, and 4, 
respectively. In each panel, heavy curves are for a nominal model with 
$t_0$ = 13 Gyr and $\Omega_B h^2$ = 0.014 and are derived from the 
foreground-corrected quadrupole-included analysis. In the two upper
panels the three light curves differ from the nominal heavy curves 
by not accounting for the faint high-latitude foreground Galactic
emission correction 
and by excluding the quadrupole from the analysis. In the two lower panels 
the three light curves differ from the nominal ones by corresponding to models
with $t_0$ = 11 Gyr and $\Omega_B h^2$ = 0.006 (panel $b)$ on the left) and 
with $t_0$ = 17 Gyr and $\Omega_B h^2$ = 0.022 (panel $d)$ on the right).}
\label{f7}
\end{figure}

\begin{figure}
\centerline{\epsfig{file=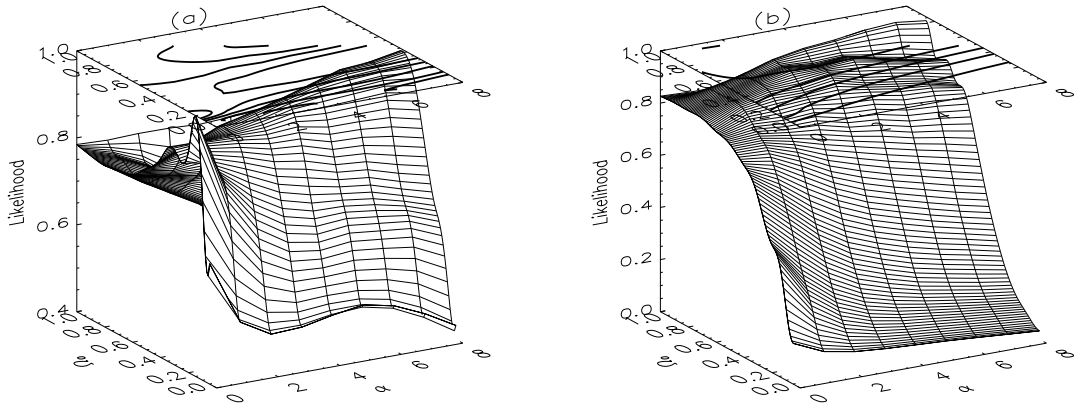,width=16.5cm}}
\caption{Likelihood functions $L(\alpha, \Omega_0)$ (arbitrarily 
normalized to unity at the highest peak) derived by marginalizing 
$L(Q_{\rm rms-PS}, \alpha, \Omega_0)$ over $Q_{\rm rms-PS}$ with a 
uniform prior. These are for the model with $t_0$ = 13 Gyr and 
$\Omega_B h^2$ = 0.014. Panel $a)$ ignores the correction for faint 
high-latitude foreground Galactic emission and excludes the quadrupole 
moment from the analysis, while panel $b)$ corrects for faint 
high-latitude foreground Galactic emission and includes the quadrupole 
moment in the analysis.}
\label{f8}
\end{figure}

\begin{figure}
\centerline{\epsfig{file=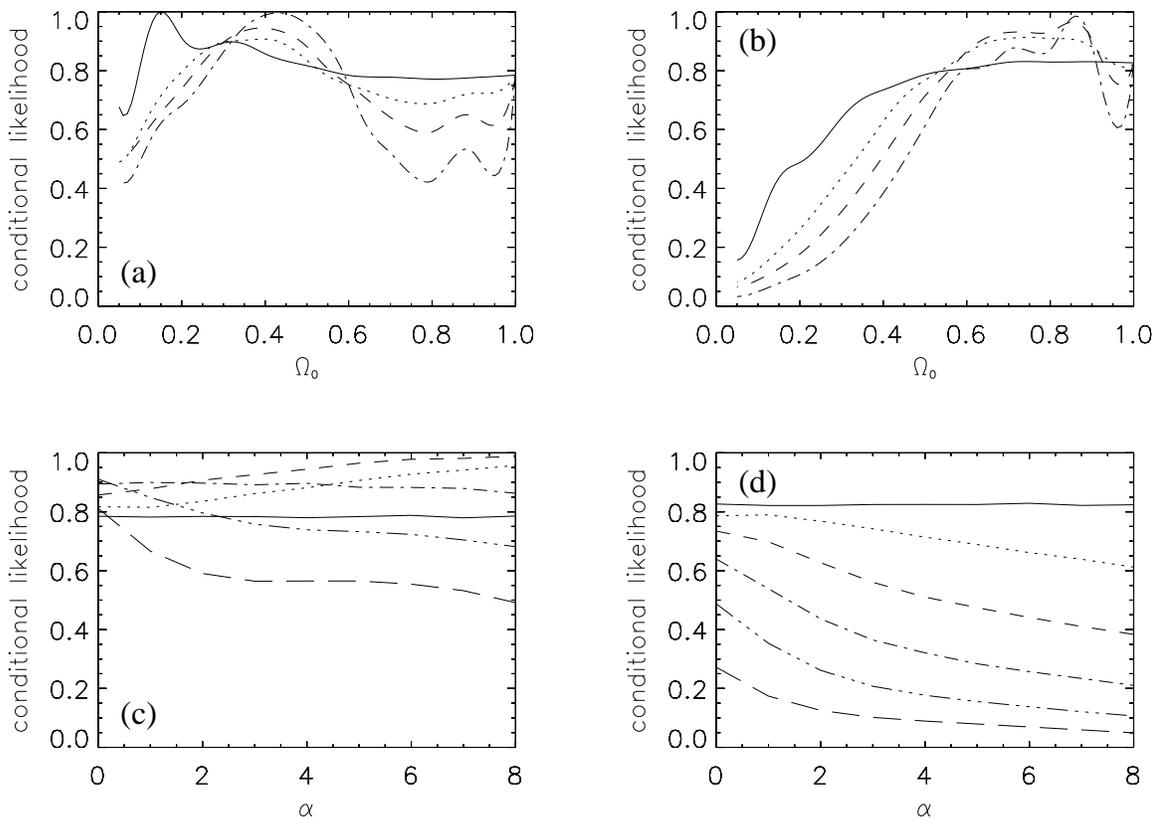,width=16.5cm}}
\caption{Conditional likelihood functions derived from $L(\alpha, \Omega_0)$ 
(arbitrarily normalized to unity at the highest peak), which is derived by 
marginalizing $L(Q_{\rm rms-PS}, \alpha, \Omega_0)$ over $Q_{\rm rms-PS}$
with a uniform prior. These are for the model with $t_0$ = 13 Gyr and 
$\Omega_B h^2$ = 0.014. Panels $a)$ and $c)$ in the left column  
ignore the correction for faint high-latitude foreground 
Galactic emission and exclude the quadrupole moment from the analysis.
Panels $b)$ and $d)$ in the right column account for the correction for 
faint high-latitude foreground Galactic emission and include the quadrupole 
moment in the analysis. The top two panels show $L(\Omega_0)$
on four constant $\alpha$ slices at $\alpha$ = 0 (solid curves), 
2 (dotted curves), 4 (dashed curves), and 8 (dot-dashed curves).
The bottom two panels show $L(\alpha)$ on six constant $\Omega_0$ 
slices at $\Omega_0$ = 1 (solid curves), 0.5 (dotted curves), 0.4 
(short dashed curves), 0.3 (dot-dashed curves), 0.2 (dash-three-dotted
curves), and 0.1 (long dashed curves).}
\label{f9}
\end{figure}

\begin{figure}
\centerline{\epsfig{file=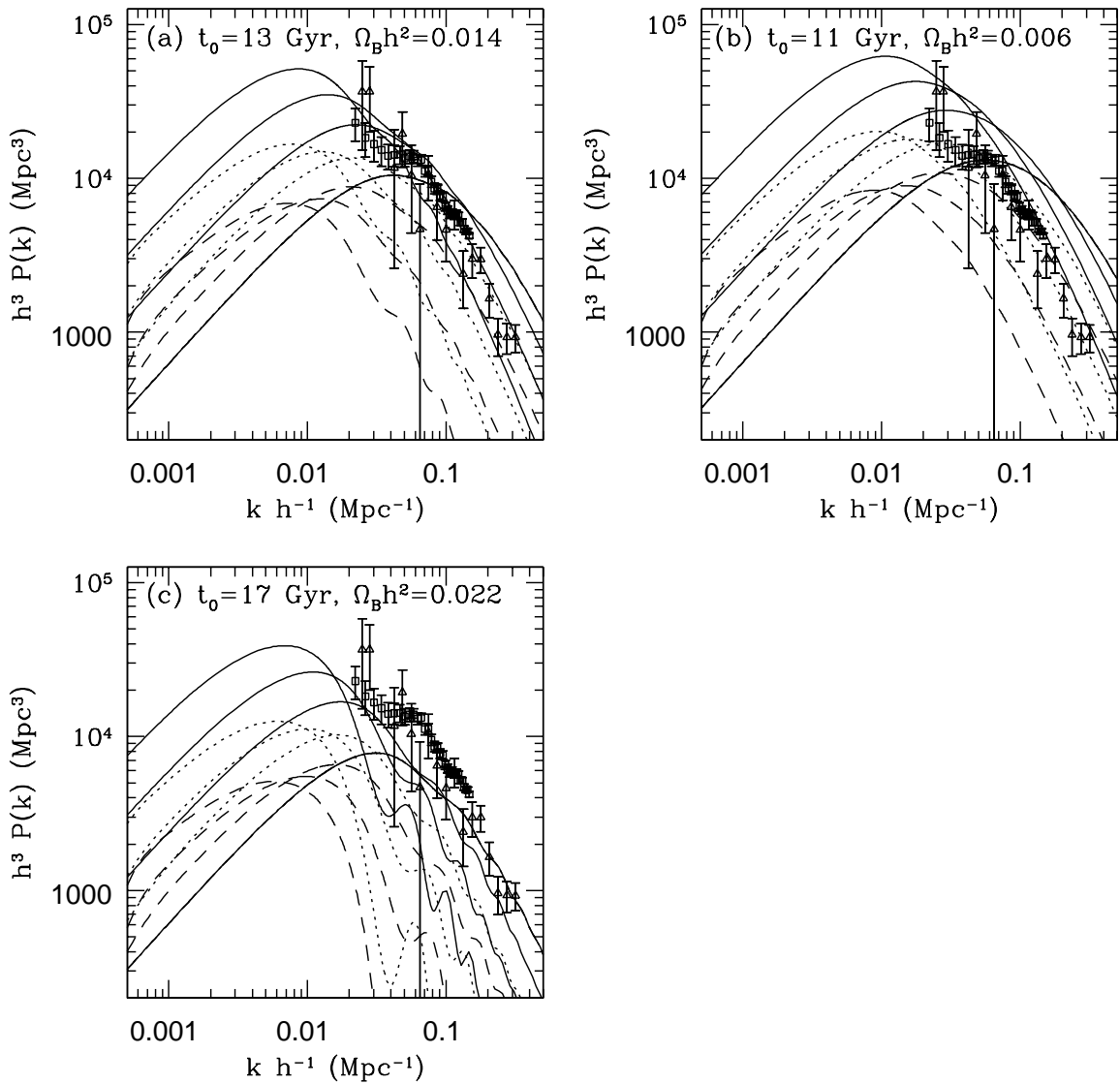,width=16.5cm}}
\caption{Fractional energy-density perturbation power spectra
$P(k)$ as a function of wavenumber $k$. These are normalized to the 
mean of the central $Q_{\rm rms-PS}$ values from the foreground-corrected
quadrupole-included analysis and the foreground-uncorrected 
quadrupole-excluded analysis (for the corresponding values of the 
cosmological parameters). Panels $a)$, $b)$, and $c)$ correspond 
to models with parameter values ($t_0$, $\Omega_B h^2$) = (13 Gyr, 0.014), 
(11 Gyr, 0.006), and (17 Gyr, 0.022), respectively. Each panel shows
12 power spectra at three values of $\alpha$ = 0 (solid curves), 2 (dotted
curves), and 4 (dashed curves) for four values of $\Omega_0$ = 1, 0.4,
0.2, and 0.1 in ascending order on the left axis of each panel. 
The triangles represent the linear real space power spectrum of the IRAS 
Point Source Catalogue Redshift Survey (PSCz) as estimated by Hamilton, 
Tegmark, \& Padmanabhan (2000). Upper limits are not plotted. 
The squares represent the redshift-space galaxy
power spectrum estimated from the 2dFGRS optical galaxy data 
(Percival et al. 2001), again on linear scales, comparison with the 
shapes of curves being valid in the case of linear bias.}
\label{f10}
\end{figure}

\end{document}